\documentclass[aps,prd,twocolumn,amsmath,amssymb,showpacs,floatfix,nofootinbib]{revtex4-1}

\usepackage{graphicx}
\usepackage{dcolumn}
\usepackage{xcolor}
\usepackage{bm}
\usepackage{natbib}
\usepackage{calligra}
\usepackage[T1]{fontenc}
\usepackage{egothic}
\usepackage[T1]{fontenc}
\newfont{\rsfsten}{rsfs10 scaled 1200}
\newfont{\rsfsseven}{rsfs10 scaled 1200}
\newfont{\rsfsfive}{rsfs10 scaled 1200}
\usepackage{epsfig}
\usepackage{units}
\usepackage[utf8]{inputenc}

\newcommand{\be}{\begin{equation}}
\newcommand{\ee}{\end{equation}}
\newcommand{\bea}{\begin{eqnarray}}
\newcommand{\eea}{\end{eqnarray}}

\def\lsim{\mathrel{\raise.3ex\hbox{$<$\kern-.75em\lower1ex\hbox{$\sim$}}}}
\def\gsim{\mathrel{\raise.3ex\hbox{$>$\kern-.75em\lower1ex\hbox{$\sim$}}}}

\begin{document}

\vskip 0.2in

\title{Orbital eccentricities in primordial black holes binaries}

\author{Ilias Cholis}
\email{icholis1@jhu.edu}
\affiliation{Department of Physics and Astronomy, The Johns Hopkins University, Baltimore, Maryland, 21218, USA}
\author{Ely D. Kovetz}
\email{ekovetz1@jhu.edu}
\affiliation{Department of Physics and Astronomy, The Johns Hopkins University, Baltimore, Maryland, 21218, USA}
\author{Yacine Ali-Ha\"{i}moud}
\email{yacine@jhu.edu}
\affiliation{Department of Physics and Astronomy, The Johns Hopkins University, Baltimore, Maryland, 21218, USA}
\author{Simeon Bird}
\affiliation{Department of Physics and Astronomy, The Johns Hopkins University, Baltimore, Maryland, 21218, USA}
\author{Marc Kamionkowski}
\affiliation{Department of Physics and Astronomy, The Johns Hopkins University, Baltimore, Maryland, 21218, USA}
\author{Julian B. Mu\~{n}oz}
\affiliation{Department of Physics and Astronomy, The Johns Hopkins University, Baltimore, Maryland, 21218, USA}
\author{Alvise Raccanelli}
\affiliation{Department of Physics and Astronomy, The Johns Hopkins University, Baltimore, Maryland, 21218, USA}

\date{\today}

\begin{abstract}

It was recently suggested that the merger of $\sim30\,M_\odot$
primordial black holes (PBHs) may provide a significant number
of events in gravitational-wave observatories over the
next decade, if they make up an appreciable
fraction of the dark matter.
Here we show that measurement of the
eccentricities of the inspiralling binary black holes can be
used to distinguish these binaries from those produced by
more traditional astrophysical mechanisms.  These PBH binaries
are formed on highly eccentric orbits and can then merge on
timescales that in some cases are years or less, retaining some eccentricity in the
last seconds before the merger.  This is to be contrasted with
massive-stellar-binary, globular-cluster, or other astrophysical
origins for binary black holes (BBHs) in which the orbits have very effectively
circularized by the time the BBH enters the observable LIGO
window.  Here we discuss the features of the gravitational-wave
signals that indicate this eccentricity and forecast the
sensitivity of LIGO and the Einstein Telescope to such effects.
We show that if PBHs make up the dark matter, then roughly one
event should have a detectable eccentricity given LIGO's
expected sensitivity and observing time of six years.  The Einstein Telescope
should see $O(10)$ such events after ten years.

\end{abstract}


\maketitle

\section{Introduction}
\label{sec:introduction}

The LIGO collaboration recently detected gravitational waves
(GWs) from the coalescence of black holes
(BHs)~\cite{Abbott:2016blz, Abbott:2016nmj}. Many viable models for the
progenitors of coalescing binary BHs have been proposed in the
literature \cite{Bird:2016dcv, Hosokawa:2015ena,
Rodriguez:2016kxx,  
Zhang:2016rli, Woosley:2016nnw, Chatterjee:2016hxc,
deMink:2016vkw, Hartwig:2016nde, Inayoshi:2016hco,
Arvanitaki:2016qwi, Rodriguez:2016avt}, all consistent with the
estimated rate (if all mergers involve black holes of similar masses) of Refs.~\cite{TheLIGOScientific:2016pea,Abbott:2016nhf}. The fact that the first detected GW
signal originated from a pair of BHs with masses $\sim
30\,M_{\odot}$ \cite{TheLIGOScientific:2016wfe} suggests that
such high-mass merger events are common enough that a 
significant sample of them will soon be obtained. 

In Ref.~\cite{Bird:2016dcv}, we suggested that if
$\sim30~M_\odot$ primordial black holes (PBH) make up the dark
matter \cite{Carr:1975qj, Carr:1974nx, Meszaros:1974tb}, then
the rate for the merger of such events is consistent with
the range inferred from the first LIGO events.  PBHs in the mass
range $20 -100\, M_{\odot}$ remain viable DM
candidates \cite{Carr:2009jm,Monroy-Rodriguez:2014ula}---lower
masses are ruled out by null microlensing searches
\cite{Allsman:2000kg, Tisserand:2006zx, Wyrzykowski:2011tr, Pooley:2008vu, Mediavilla:2009um} and
pulsar-timing-array searches \cite{Bugaev:2010bb}, and higher
masses by the dynamics of wide stellar binaries
\cite{Yoo:2003fr,Quinn:2009zg,Monroy-Rodriguez:2014ula}.
Ref.~\cite{Ricotti:2007au} inferred a strong constraint to the PBH abundance in the $\sim30\,M_\odot$ mass range from the cosmic microwave background spectrum and anisotropies. Ref.~\cite{Brandt:2016aco} discussed an interesting
constraint, from a weakly bound stellar cluster, to dark matter
in this mass range. 
Given the caveats associated with both results, however, neither constraint is 
strong enough to robustly exclude the possibility of PBH dark matter. In particular, 
it is conceivable, given the large number of GW events
likely to be detected within the next decade, that some may be
PBH binaries, even if PBHs are only a sub-dominant
constituent of dark matter \cite{Bird:2016dcv}.

Several ideas have already been proposed to test the possibility
that $\sim30~M_\odot$ PBHs make up all or part of the dark
matter.  For example, merging PBHs are likely to reside in
lower-mass halos \cite{Bird:2016dcv}, so the cross-correlation
of future GW event-location maps with galaxy catalogues may test
the PBH-progenitor model \cite{Raccanelli:2016cud, Namikawa:2016edr}.  The
scenario may also be tested by seeking fast-radio-burst echoes
induced by strong gravitational lensing by PBHs \cite{Munoz:2016tmg}, or additional
work along the lines of the CMB, wide-binary, or stellar-cluster
probes discussed above.

In this paper we explore the possibility to distinguish directly
with gravitational-wave measurements whether PBH mergers
contribute some of the gravitational-wave events that will be
observed.  In particular, we show here that some PBH binaries through their evolution
will have eccentricities that are large enough to be detectable even in
the final stages of inspiral before the merger.  This contrasts
with the expectation for most other progenitor models, in which
the orbits will have very effectively circularized by the time
they reach the observable gravitational-wave window.

The velocity dispersion of dark-matter PBHs in the galactic
halos in which they reside depends on the size of the DM
halo. For example, PBHs in Milky-Way sized halos
typically have relative velocities of $O(10^{2})$ km~s$^{-1}$, while in 
halos as small as $10^{3}$ $M_{\odot}$, those velocities are
suppressed to $O(0.1)$ km~s$^{-1}$.  PBH binaries are formed
when two PBHs pass so close that they emit enough energy to GW
radiation to form a bound pair.  Higher relative velocities
result in tighter binaries with highly eccentric (nearly
parabolic) orbits. As the PBHs coalesce, the orbit will
gradually circularize. The characteristic merger time, though,
can vary significantly, leading to a wide range of residual
orbital eccentricities, as different merger events enter the
frequency band of a GW detector.  The signature of higher
eccentricities is the observation of the higher harmonics they
produce in the gravitational-wave signal.  Here we focus in
particular on the possibility pf observing these higher harmonics
with advanced LIGO \cite{Aasi:2013wya} and the planned Einstein
Telescope (ET) \cite{Sathyaprakash:2011bh}.  We find that the former
is likely to see $O(1)$ and the latter  $O(10)$ such eccentric
mergers of $\sim 30 M_{\odot}$ BHs over the span of ten years,
providing an important test of the PBH scenario.

Before proceeding, we note that
Refs.~\cite{Clesse:2016vqa,Sasaki:2016jop} have argued for
early-Universe mechanisms in which PBHs would be formed in
binaries.  The mergers of such primordial PBH binaries would
occur only after the primordial binaries have had considerable
opportunity to circularize.  The large-eccentricity signature of
PBHs that we discuss here thus applies only to the scenario
envisaged in Ref.~\cite{Bird:2016dcv}, in which PBH binaries form
late in the Universe via GW radiation.

This paper is organized as follows. In
Section~\ref{sec:Preliminaries} we  review the two-body GW
mechanism for binary capture, as well as the equations for the
evolution  of the orbital properties during the subsequent
merger. In Section~\ref{sec:PBHs_Binaries}  we analyze the case
of PBH binaries and calculate their initial orbital properties
and their evolution, including the time from formation until merger 
and their final eccentricities. We discuss also the effects that a third body 
may have on the existing PBH binaries.  
In Section~\ref{sec:Detectability} we
address the detectability of the high-eccentricity PBH-merger
events by current and future GW observatories. We first estimate
the expected event rate for eccentric events and compare it to
other progenitor models for the formation of BBHs, that in particular occur at globular clusters
or in environments near supermassive BHs. Furthermore we consider the resulting
gravitational wave modes and their observational consequences.  We conclude in
Section~\ref{sec:Conclusions}.
               
\section{Preliminaries}
\label{sec:Preliminaries}
In this Section we provide the formulas that are used to derive
our results in Section~\ref{sec:PBHs_Binaries}. 
Unless explicitly specified we use geometric units with $G = c = 1$.

\subsection{Two-body binary capture}

We consider the formation of a binary from two BHs with masses $m_{1}$ 
and $m_{2}$ and a mass ratio defined as,
\begin{eqnarray}
\eta(m_{1}, m_{2}) = \frac{m_{1}m_{2}}{(m_{1}+m_{2})^{2}}
\equiv\frac{m_{1}m_{2}}{m_{\rm tot}^{2}}.
\label{eq:eta}
\end{eqnarray}
As the two BHs approach each other with a relative velocity $w$, they emit 
gravitational waves whose power peaks at $r_{p}$, 
the distance of closest approach, which is related to the impact parameter $b$ by
 \cite{O'Leary:2008xt}
\begin{eqnarray}
&b&= \frac{\sqrt{2 m_{\rm tot} r_p}}{w} \left(1 + \frac{r_p w^2}{2 m_{\rm tot}}\right)^{1/2} \textrm{or to first order in $w/c$}, \nonumber \\
&r_{p}&(m_{1}, m_{2}, w, b) \simeq \frac{b^{2} w^{2}}{2 m_{\rm tot}} 
\left( 1 - \frac{b^{2} w^{4}}{4 m_{\rm tot}^{2}}  \right).
\label{eq:pericenter}
\end{eqnarray}
In order for the two BHs to create a binary after the close
encounter, enough energy has to be radiated via GWs. For such
systems, the {\it final} energy right after formation is given
by \cite{Peters:1963ux, 1977ApJ...216..610T}
\begin{eqnarray}
E_{f}(m_{1}, m_{2}, w, b) = \frac{m_{\rm tot}\eta w^{2}}{2} 
- \frac{85 \pi }{12\sqrt{2}}\frac{\eta^{2} m_{\rm tot}^{9/2}}{r_{p}^{7/2}},~~~~
\label{eq:Ef}
\end{eqnarray}
where the second term is the energy released in GWs, 
and the maximum impact parameter to form a bound binary is
\cite{O'Leary:2008xt},
\begin{eqnarray}
b_{\rm max}(m_{1}, m_{2}, w) = \left( \frac{340 \pi}{3} \right)^{1/7} 
\frac{m_{\rm tot} \eta^{1/7}}{w^{9/7}}.
\label{eq:bmax}
\end{eqnarray}
Once formed, the binary's initial semi-major axis is
\begin{eqnarray}
a_{0}(m_{1}, m_{2}, w, b) = - \frac{m_{\rm tot}^{2} \eta }{2 E_{f}},
\label{eq:semimajor_axis}
\end{eqnarray}
while its initial eccentricity is
\begin{eqnarray}
e_{0}(m_{1}, m_{2}, w, b) = \sqrt{1+ 2\frac{E_{f} b^{2}w^{2}}{m_{\rm tot}^{3} \eta}},
\label{eq:e0}
\end{eqnarray}
and the initial Keplerian-orbit pericenter distance is given by
$r_{p_{0}} = a_{0} (1 - e_{0})$, or in its dimensionless parametrization; $\rho_{p_{0}} = r_{p_{0}}/m_{\rm tot}$.

It will be useful in what follows to rewrite these expressions as a function of $b/b_{\max}(w)$:
\begin{eqnarray}
E_f &=& - \frac12 m_{\rm tot} \eta w^2 \left[\left(\frac{b_{\max}}{b}\right)^7 - 1\right],\\
a_0 &=& m_{\rm tot} w^{-2} \left[\left(\frac{b_{\max}}{b}\right)^7 - 1\right]^{-1}, \label{eq:a0_b}\\
1 - e_0^2 &=& \left(\frac{340 \pi \eta}{3} \right)^{2/7} w^{10/7} \left(\frac{b}{b_{\max}}\right)^2\left[\left(\frac{b_{\max}}{b}\right)^7 - 1\right].~~~~ \label{eq:e0_b}
\end{eqnarray}

\subsection{Merger timescale and orbital evolution}

For a binary with initial eccentricity $e_{0} \simeq 1$ and semi-major
axis $a_{0}$, the time it takes it to merge is
given by \cite{Peters:1964zz},
\begin{eqnarray}
\tau_{m}(m_{1}, m_{2}, e_{0}, a_{0}) &=& \frac{15}{19} 
\left( \frac{304}{425} \right)^{\frac{3480}{2299}} 
\frac{m_{\rm tot} \rho_{p_{0}}^{4} }{4 \eta}\times  \nonumber \\ 
&&\times\int_{e_{\rm LSO}}^{e_{0}} de \; e^{29/19} 
\frac{\left( 1 + \frac{121}{304} e^{3} \right)^{\frac{1181}{2299}}}{\left(1-e^{2}\right)^{3/2}}.
\nonumber \\ 
\label{eq:tau_m_1}
\end{eqnarray}
Given that $e_{0}$ and $a_{0}$ can be traced back to the initial
relative velocity and impact parameter of the BHs,
Eq.~\ref{eq:tau_m_1} can be recast as,
\begin{eqnarray}
\tau_{m}(m_{1}, m_{2}, w, b)& =& 
\frac{3}{85}\frac{a_{0}^{4}}{m_{\rm tot}^{3} \eta}\left(1-e_{0}^{2}\right)^{7/2}.
\label{eq:tau_m_2}
\end{eqnarray}
Thus in our analysis of PBH binaries, we will simply need to
simulate the distributions of $w$ and $b$ for a given choice of
$m_{1}$ and $m_{2}$. This will be done in the next Section.

In order to track the evolution of the orbital eccentricity, we
evolve the BH binaries until their final eccentricity $e$, given
an initial value $e_{0}$ and some initial pericenter distance $r_{p_{0}}$
and a final pericenter distance $r_{p_{f}}$. 
The semi-major axis and eccentricity decrease due to angular momentum and energy loss through gravitational-wave radiation. The resulting coupled ordinary differential equations can be rewritten as 
\begin{eqnarray}
r_p \frac{de}{d r_p} = e (1 + e) \frac{304 +121 e^{2}}{192 - 112 e + 168 e^{2} +47 e^{3}}.
\label{eq:dedchi}
\end{eqnarray}
This equation can be integrated analytically and the solution takes the form of an algebraic equation \cite{Peters:1964zz}
\begin{eqnarray}
\frac{r_p}{r_{p_0}} = \left(\frac{e}{e_0}\right)^{12/19}\frac{\mathcal{F}(e)}{\mathcal{F}(e_0)}, \label{eq:rp_of_e}
\end{eqnarray}
with 
\begin{eqnarray}
\mathcal{F}(e) \equiv (1 + e)^{-1} \left(1 + \frac{121}{304} e^2 \right)^{\frac{870}{2299}}.
\end{eqnarray}
We wish to find the eccentricity $e_f$ when the pericenter reaches some final value $r_{p_f}$. Eq.~\eqref{eq:rp_of_e} is implicit in $e_f$,
\begin{eqnarray}
e_f = e_0 \left( \frac{r_{p_f}}{r_{p_0}} \frac{\mathcal{F}(e_f)}{\mathcal{F}(e_0)} \right)^{19/12}.
\end{eqnarray}
However, it can be solved very efficiently through the following iterative method: we set $e_f^{(0)} \equiv 0$ and 
\begin{eqnarray}
e_f^{(i+1)} \equiv  e_0 \left( \frac{r_{p_f}}{r_{p_0}} \frac{\mathcal{F}(e_f^{(i)})}{\mathcal{F}(e_0)} \right)^{19/12}.
\end{eqnarray}
We iterate until reaching a 1\% convergence. We have checked that the solution obtained this way matches that obtained from explicitly solving Eq.~\eqref{eq:dedchi}.

\section{The Case of DM PBH Binaries}
\label{sec:PBHs_Binaries}

\subsection{Assumed distributions of PBHs in DM halos}
\label{sec:PBHdistr}

In deriving the distribution functions for PBHs in DM halos, we
have to make some assumptions regarding the velocity
distribution of the PBHs. Following \cite{Bird:2016dcv}, we take
a Maxwell-Boltzmann distribution,
\begin{eqnarray}
P_{v_{\rm PBH}}(v) = F_{0}^{-1} v^{2} \left (e^{-v^{2}/v_{\rm DM}^{2}} - e^{-v_{\rm vir}^{2}/v_{\rm DM}^{2}} \right), 
\label{eq:PvPBH}
\end{eqnarray}
with $F_{0} = 4 \pi \int_{0}^{v_{\rm vir}} v^{2} \left (e^{-v^{2}/v_{\rm DM}^{2}} - e^{-v_{\rm vir}^{2}/v_{\rm DM}^{2}} \right) $
and where the velocity dispersion $v_{\rm DM}$ and the virial
velocity $v_{\rm vir}$ depend on the DM halo mass. For a
$10^{12} M_{\odot}/h$ object for instance, we have $v_{\rm DM} =
166$ and $v_{\rm vir} = 200$ km~s$^{-1}$.  The typical relative
velocity of PBHs in a $10^{6}$ ($10^{9}$, $10^{12}$)
$M_{\odot}/h$ DM halo is $w = \sqrt{2}v=2$ ($\times 10^{1}, \times 10^{2}$)
km~s$^{-1}$, and for $m_{1}=m_{2}=30 M_{\odot}$, we get $b_{\rm
max} = 5.1$ (0.26, $1.3 \times 10^{-2}$) au.

As for the impact parameter, we assume that $b^{2}$ is uniformly distributed between 0 and $b_{\rm max}^{2}(30 M_{\odot}, 30 M_{\odot}, w)$.

Using these distributions, we simulate the formation of $10^{6}$
binary PBHs of $30\,M_{\odot}$.  Their rate of formation in a DM
halo of a given mass is described in Eqs.~(8) and (9) of
Ref.~\cite{Bird:2016dcv}; where it was shown that the total rate of
merging PBH binaries, overlaps with the range of 2-53 per Gpc$^{3}$ 
per yr quoted by LIGO \cite{Abbott:2016nhf}. 
Recently, the rate of $\sim30\,M_\odot$ merging black holes was
reevaluated to be between $0.5-11.5$ mergers per Gpc$^{3}$ per year 
\cite{TheLIGOScientific:2016pea}. In the remainder of the paper we will normalize the rate of PBH 
to this updated observed LIGO rate. 

\subsection{Initial inspiral properties}
\label{sec:Inital}
\begin{figure}[h!]
\begin{centering}
\includegraphics[width=\columnwidth]{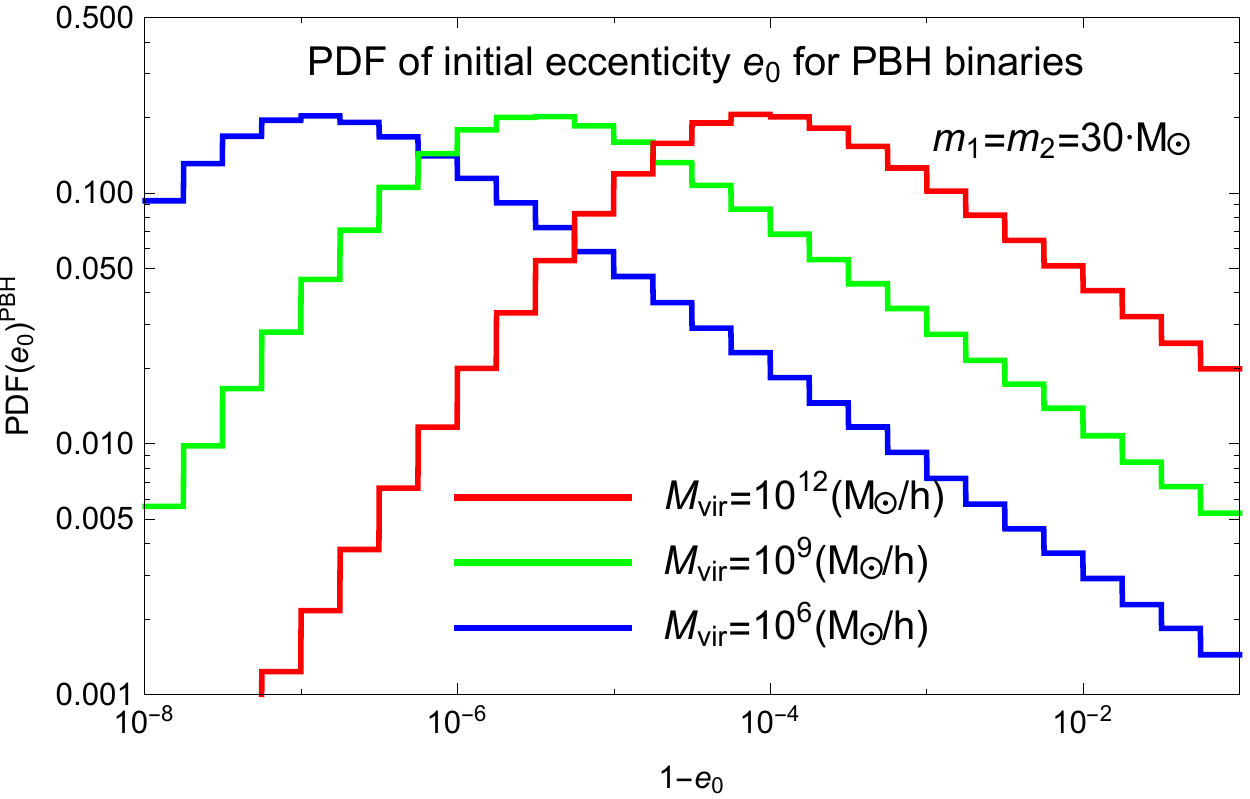}
\end{centering}
\caption{The distributions PDF$(e_{0})$ of eccentricities at
formation, for PBH binaries residing in DM halos of three
different sizes, based on $10^6$ simulations. In \textit{all
cases} the PBH binaries have highly eccentric orbits, with their
respective distributions peaking as expected (see text for
details).}
\label{fig:initial_eccentricities}
\end{figure}

The probability distribution functions (PDFs) for the
eccentricity at binary formation is shown in
Fig.~\ref{fig:initial_eccentricities}, where we present our
results for binaries residing in three different sizes of DM
halos, i.e.\ virial masses of $10^{6}$, $10^{9}$ and $10^{12}$
$M_{\odot}/h$.  We find that in all cases the eccentricity
of the BHs is close to unity at formation regardless of the mass
of the host halo. In fact, a heuristic argument can be made for
where the peak of the PDF of the eccentricity distribution, $(1 -
e_{0})^{\textrm{peak}}$ should be.  The $(1 -
e_{0})^{\textrm{peak}}$ $= r_{p_{0}}/a_{0}$ which from
Eqs.~(\ref{eq:pericenter}) (at 0-th order)
and~(\ref{eq:semimajor_axis}) is -$b^{2} w^{2} E_{f}$$/(m_{\textrm{tot}}^{3} \eta)$.  Approximating $b$ with
$b_{\textrm{max}}$ (given by Eq.~(\ref{eq:bmax})), we get $(1 -
e_{0})^{\textrm{peak}}$ $\simeq - (340 \pi/3)^{2/7}$ $w^{-4/7}
\eta^{-5/7} E_{f}/m_{\textrm{tot}}$. Since the initial
energy of the system is $E_{i} = \eta m_{\textrm{tot}} w^{2}/2$,
and defining $\xi = |E_{f}|/E_{i}$, this gives $(1 -
e_{0})^{\textrm{peak}}$ $\simeq 2.6  \, \xi \, \eta^{2/7} \,
(w/c)^{10/7}$. Matching to the peaks in the simulations, shown in Fig.~\ref{fig:initial_eccentricities}, 
they fall at  $\xi \simeq 1$ for $w=2$ ($\times 10^{1}, \times 10^{2}$)
km~s$^{-1}$. As we discuss below, binary BHs created dynamically
in globular clusters have flatter eccentricity distributions at
formation.

\begin{figure}
\begin{centering}
\includegraphics[width=\columnwidth]{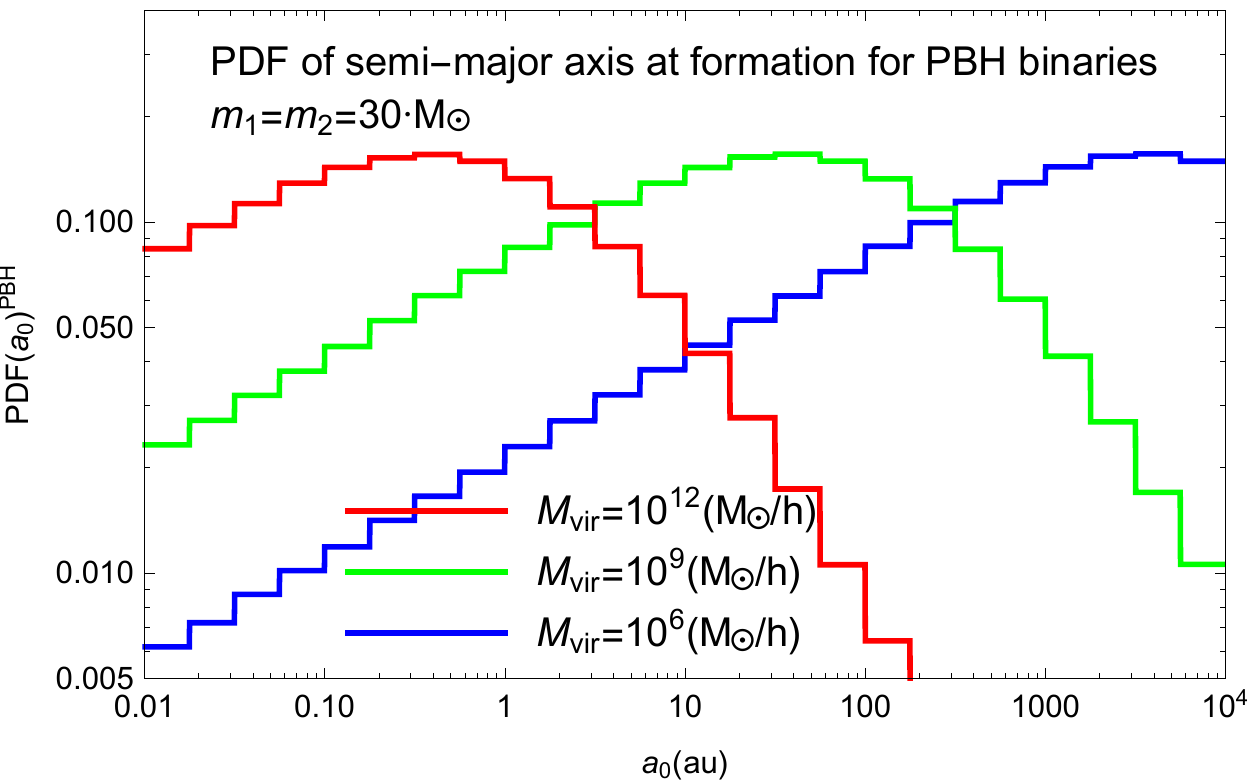}
\end{centering}
\caption{The distribution PDF$(a_{0})$ of the semi-major axis $a_{0}$ at formation. 
We consider PBH binaries residing in DM halos of three different
sizes. The formed binaries in larger halos 
 have smaller distance separation. Even for PBH  binaries in
$10^{6}$ $M_{\odot}$ halos, $a_{0}$ only goes up to $O(10^{4})$
au. In all cases we used $10^{6}$ simulated BH binaries.}
\label{fig:initial_a0}
\end{figure}

\begin{figure}
\begin{centering}
\includegraphics[width=\columnwidth]{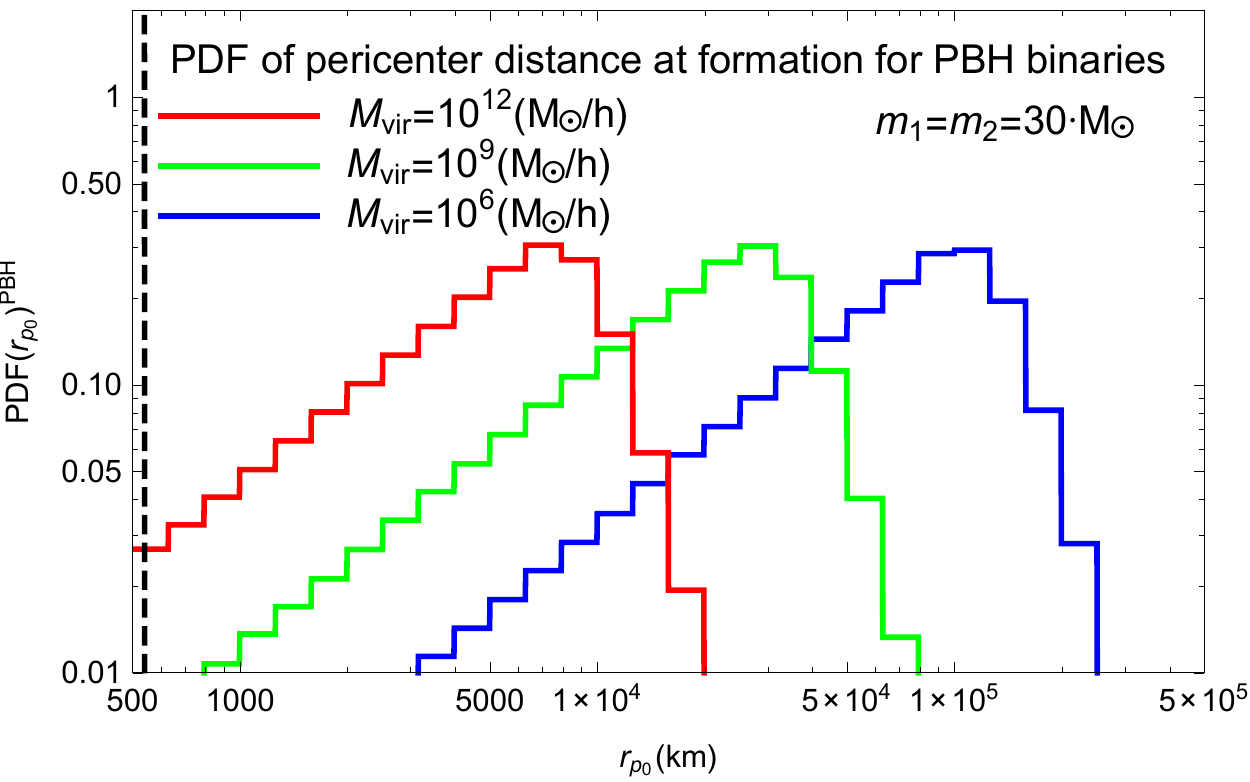}
\end{centering}
\caption{The distribution of the pericenter distance $r_{p_{0}}$
at binary formation PDF$(r_{p_{0}})$. As in
Figs.~\ref{fig:initial_eccentricities} and~\ref{fig:initial_a0},
we show binary PBHs residing at three different DM halo
sizes. Some of the formed PBH binaries at Milky Way sized and
dwarf galaxy sized  halos have pericenter distances at formation
very close to (and even in some cases less than) the last stable
orbit distance of 6 $R_{Sch}$ (black dashed line), suggesting direct
plunges. In all cases we used $10^{6}$ simulated BH binaries.}
\label{fig:initial_rp0}
\end{figure}

\begin{figure}
\begin{centering}
\includegraphics[width=\columnwidth]{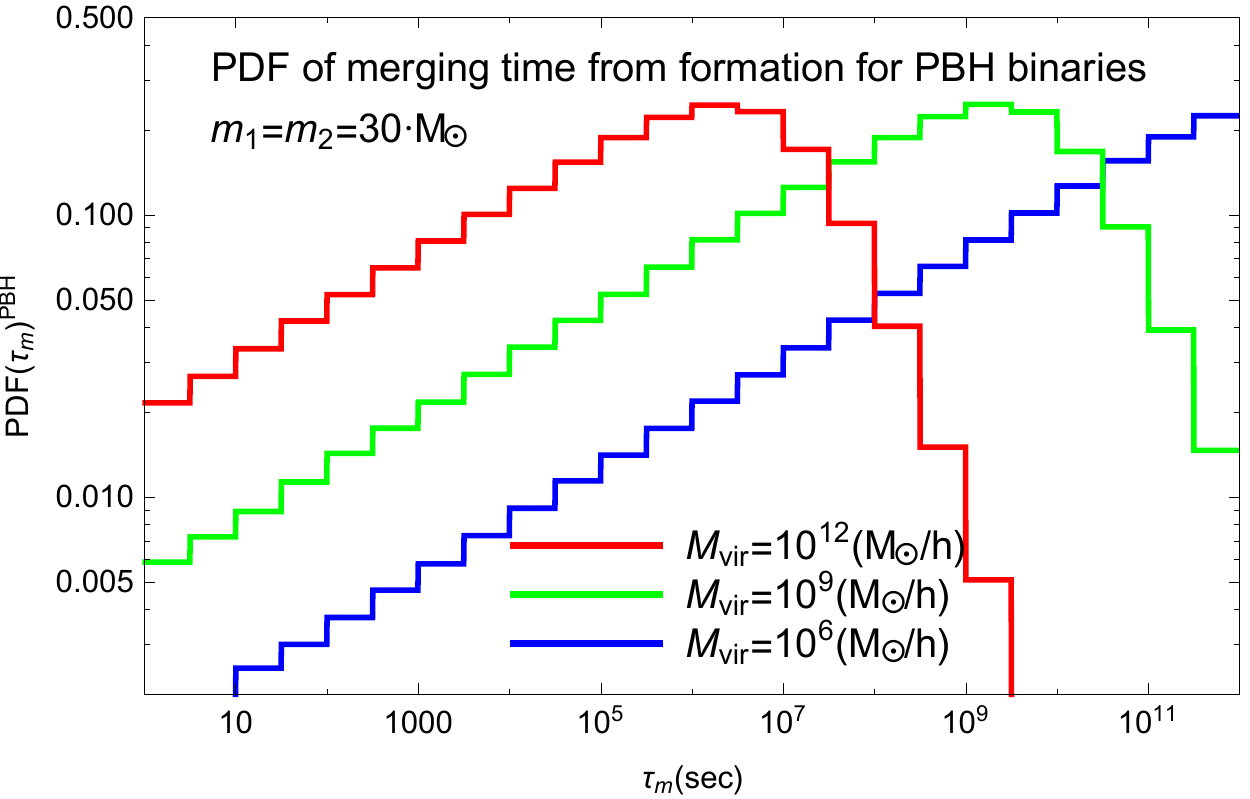}
\end{centering}
\caption{The distribution PDF$(\tau_{m})$ of time between binary formation and merger, 
considering host DM halos of three different sizes. Binaries of PBHs in Milky Way-like 
halos tend to merge within months ($\lsim 10^{7}$s) of their formation, with a significant portion merging 
within hours ($\lsim 10^{4}$s). Even for PBH binaries in $10^{6}$ $M_{\odot}$ halos, there is a few percent
chance that they merge within hours of their formation.}
\label{fig:merger_times}
\end{figure}

The PDFs for the semi-major axis, the initial pericenter
distance and the time $\tau_{m}$ from formation until merger are
shown in Figs.~\ref{fig:initial_a0},~\ref{fig:initial_rp0} and
\ref{fig:merger_times}, respectively. Here again, we can derive simple scalings analytically. From Eqs.~\eqref{eq:a0_b} and \eqref{eq:tau_m_2} we get, for $m_{\rm tot} = 60 M_{\odot}$, 
\begin{eqnarray}
a_0 &\sim& m_{\rm tot} v_{\rm DM}^{-2} \sim 100 ~\textrm{au}~ (v_{\rm DM}/20~\textrm{km s}^{-1})^{-2},\\
r_{p_0} &=& a_0 (1 - e_0) \sim 2\times 10^4 ~\textrm{km}~ (v_{\rm DM}/20~\textrm{km s}^{-1})^{-4/7},~~~~\\
\tau_m &\sim& 10^{9} \textrm{s} ~ (v_{\rm DM}/20~\textrm{km s}^{-1})^{-3}.
\end{eqnarray}

These quantities depend strongly on the DM halo the binaries
form in. Due to the difference in velocity dispersions between
halos of different masses, the characteristic time it takes a BH
binary to merge can range from minutes-hours for $M_{\rm vir}
\simeq 10^{12}\,M_{\odot}/h$ up to 100 kyrs for $M_{\rm vir} \simeq
10^{6}\,M_{\odot}/h$. In environments with high relative
velocities the capture rate is much smaller, as was shown
in Ref.~\cite{Bird:2016dcv}, but when such captures  
occur they result in tight binaries that merge much faster.
For any choice of parameters, the merger timescales for PBH
binaries are always small enough that the evolution of the DM
halo during the coalescence can be neglected. Therefore, the
merger rate can be taken to equal the rate of PBH binaries
formation. This is generally not the case for binaries formed in
other astrophysical environments where the initial separations
are typically larger and the merger timescales can be as large
as Gyrs~\cite{Kulkarni:1993fr, Kalogera:2003pk,
Kalogera:2006uj, Vanbeveren:2008sj, Dominik:2013tma,
Mennekens:2013dja, Dominik:2014yma, Mandel:2015qlu,
Chatterjee:2016hxc}.  This difference in merger timescales is an
essential characteristic in discriminating between PBH and other
progenitors.

If the impact parameter is small enough, the pericenter at first passage can be less than $6 R_{\rm Sch}$, the radius of the innermost stable orbit ($R_{\rm Sch}$ is the Schwarzschild radius of the 30 $M_{\odot}$ PBHs). In this case the BHs practically collide (plunge) at first encounter. Such events would also produce
strong GW signals. However, these would not follow the same
waveform evolution as typical coalescence events currently
searched for by LIGO. Based on Eq.~\eqref{eq:pericenter} to lowest order in $w$, we see that this corresponds to a minimum impact parameter 
\begin{eqnarray}
b_{\min}(w) &=& \sqrt{12} m_{\rm tot} w^{-1} \nonumber\\
&=& \sqrt{12} \left(\frac{3}{340 \pi \eta}\right)^{1/7} w^{2/7} b_{\max}(w). \label{eq:bmin}
\end{eqnarray}
The fraction of binary formation event that are direct plunges is therefore of order 
\begin{eqnarray}
(b_{\min}/b_{\max})^2 &\sim& 12 \left(\frac{3}{340 \pi \eta}\right)^{1/7} v_{\rm DM}^{4/7} \nonumber\\
&\sim& 1 \% ~(v_{\rm DM} / 20~ \textrm{km s}^{-1})^{4/7}.
\end{eqnarray}
Numerically, we find that for PBHs of 30 $M_{\odot}$ 
residing in $10^{6}$ ($10^{9}$, $10^{12}$) $M_{\odot}/h$ 0.3$\%$ (1.3$\%$, 4$\%$) of the interactions for which $E_{f}<0$ fall in that category. 

Figs.~\ref{fig:initial_eccentricities}-\ref{fig:merger_times} do
include those plunges. To search for
such events, a better understanding of the expected signals is
needed, most likely through numerical-relativity simulations. In
the remainder of the paper we exclude such plunge events when
referring to BH binary mergers.

\subsection{Final eccentricities}

\begin{figure}
\begin{centering}
\includegraphics[width=\columnwidth]{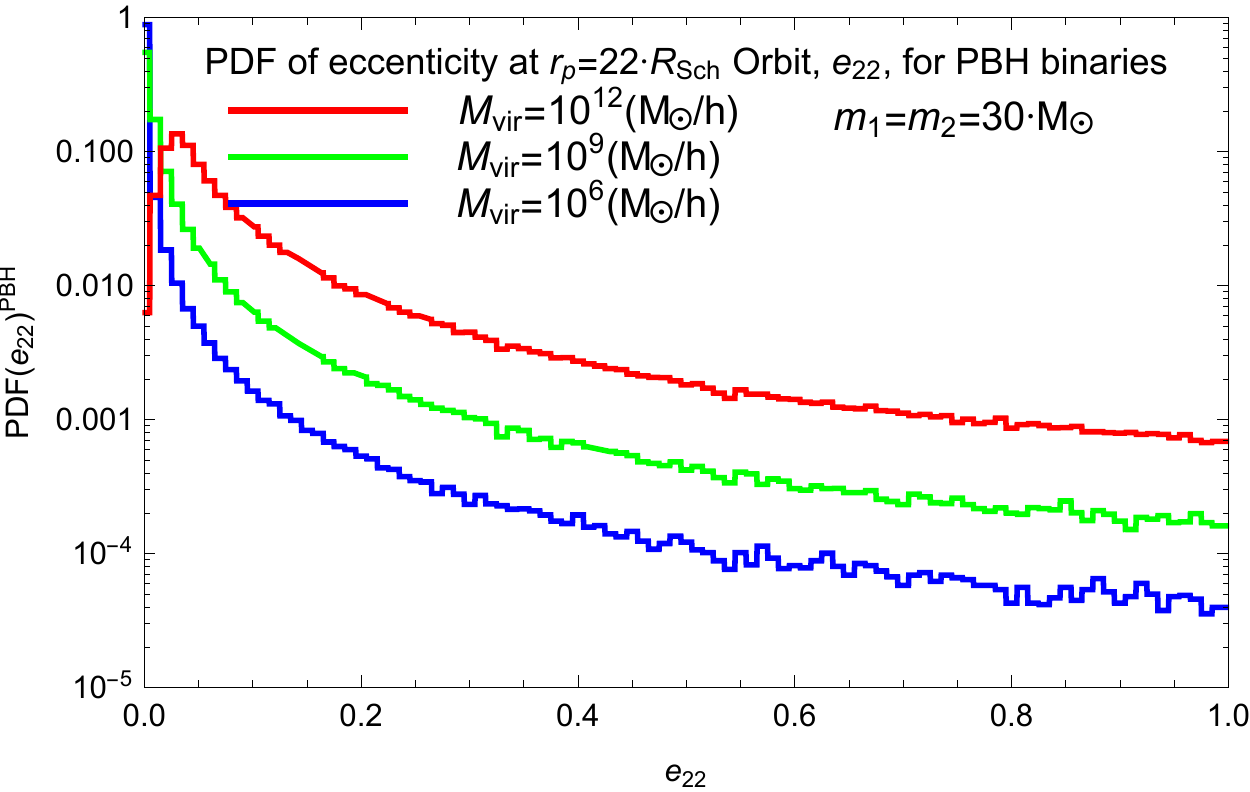}
\includegraphics[width=\columnwidth]{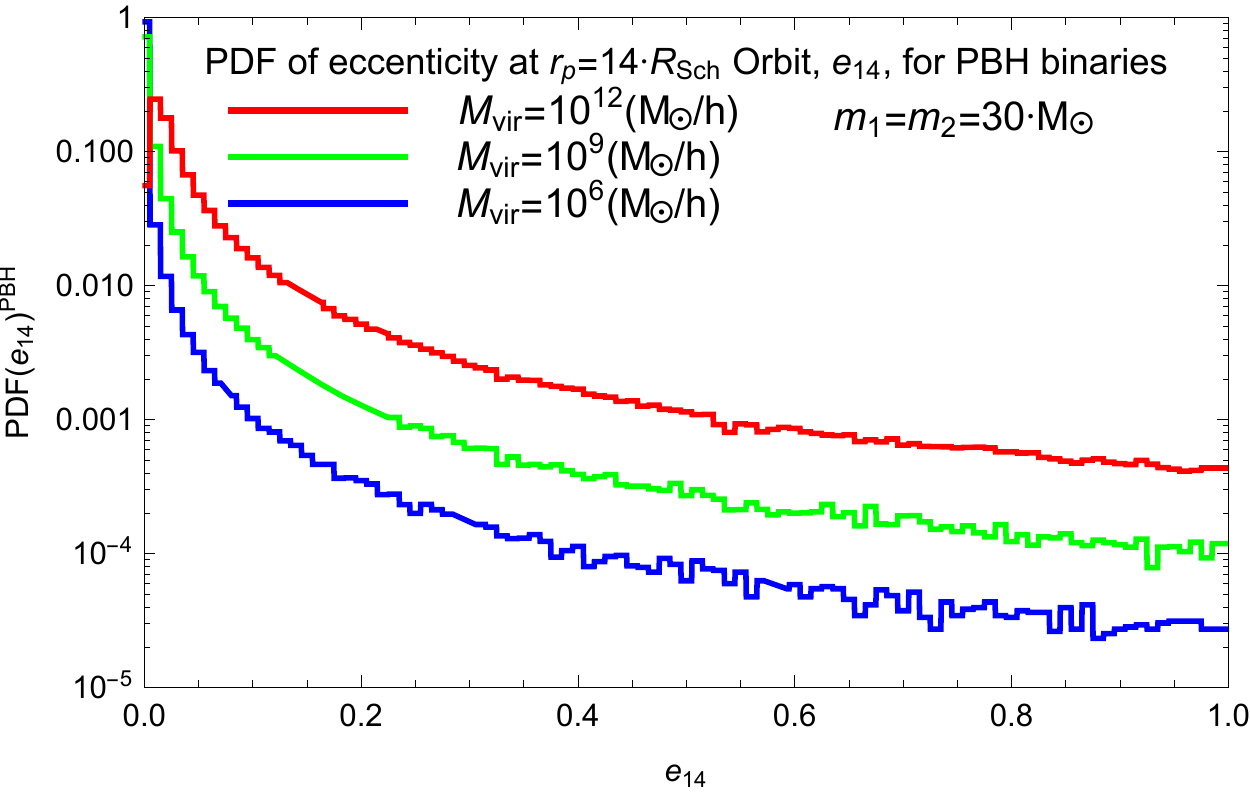}\\
\includegraphics[width=\columnwidth]{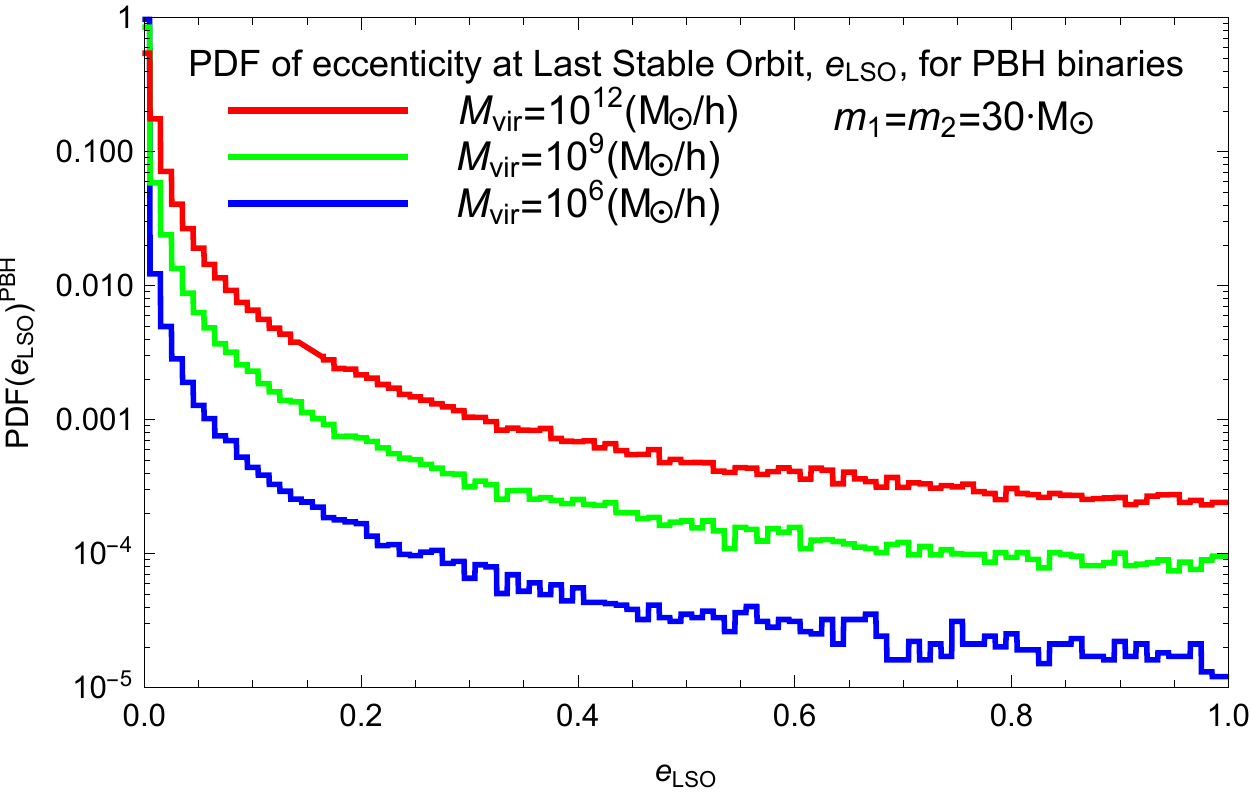}
\end{centering}
\caption{The distribution of eccentricities at three different
pericenter distances for PBHs.  The eccentricities $e_{22}$,
$e_{14}$ and $e_{\rm LSO}$ refer to the orbital eccentricity
when the pericenter distance $r_{p}$ is 22, 14 and 6 $R_{\rm
Sch}$, respectively, near enough 
to enter the LIGO and ET observed frequency bands. As before, we show results for three
different host-halo masses. PBH binaries in Milky Way-sized halos, although they have a much lower 
formation rate, retain higher eccentricities 
up to the latest stages (due to the smaller impact parameter
required for their formation). PBH binaries residing at $10^{6}$ $M_{\odot}/h$ 
have a $\sim0.1\%$ chance to remain in an eccentric orbit up to the late stages.}
\label{fig:eLSO}
\end{figure}

Following the binary evolution until the last stable orbit of $6
R_{\rm Sch}$,
we derive their final eccentricity distribution in
Fig.~\ref{fig:eLSO}. We show the distributions of eccentricities
for three different pericenter distances. These are at 22, 14
and 6 $R_{\rm Sch}$. 
We choose 22 $R_{\rm Sch}$ as this is the distance at which 
we estimate the binary to enter the LIGO band of
observations: a pair of 30 $M_{\odot}$ BHs will merge at
an orbital frequency of $\simeq 35$Hz ($\simeq 70$ Hz for the
quadrupole mode). LIGO at final design will be able to detect
down to orbital frequencies of 5 Hz (or quadrupole frequencies of 10
Hz). Thus LIGO with enough sensitivity can observe such a
system's orbital period evolution out to a factor of 7. Using
Kepler's third law of motion, this results in a semi-major axis
evolution by a factor of $7^{2/3} = 3.7$. For fixed eccentricity,
the pericenter distance will evolve by the same factor; 
i.e from $3.7 \times 6 = 22$ $R_{\rm Sch}$ to 6 $R_{\rm
Sch}$. Realistically, since the eccentricity will also be
reduced, the evolution in the pericenter distance is
smaller. The value of 14 $R_{\rm Sch}$ is thus also presented 
as an intermediate case.

\begin{figure}
\begin{centering}
\includegraphics[width=\columnwidth]{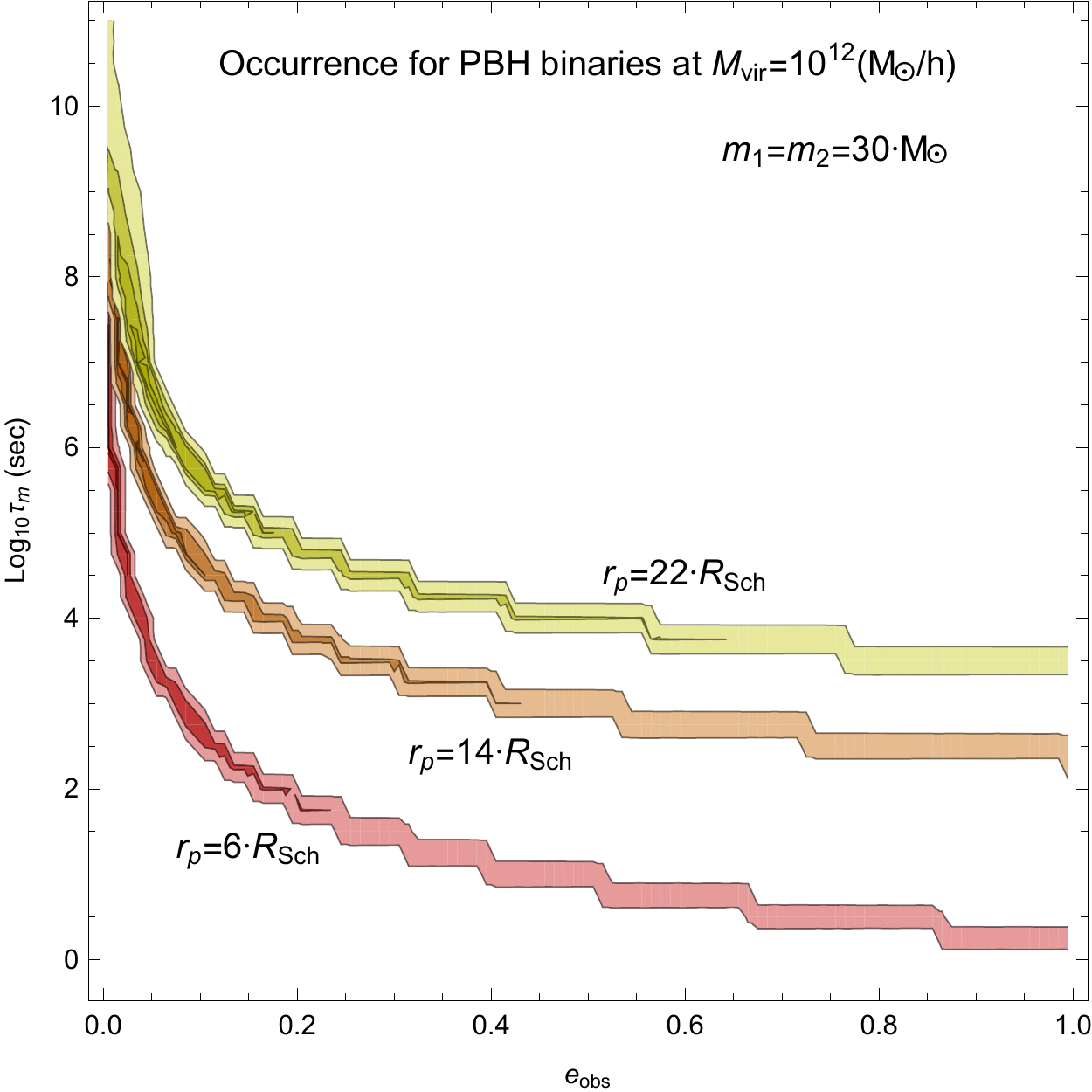}
\end{centering}
\caption{The distribution of time $\tau_{m}$ between binary formation and merger versus the 
$e_{22}$, $e_{14}$ and $e_{\rm LSO}$ eccentricities, for binaries of PBHs residing in 
$10^{12}\,M_{\odot}/h$. Color brightness indicates the occurrence of binaries. We show 
regions including $33\%$, $68\%$, $95\%$ and $99.5\%$ of the $10^{6}$ simulated BH binaries. 
Binaries with merging times of more than $10$ yrs do not retain eccentricities larger than $0.05$ 
by the time they enter the LIGO band.}
\label{fig:merger_times_VS_eLSO}
\end{figure}

Fig.~\ref{fig:eLSO} in combination with
Fig.~\ref{fig:merger_times} suggests that binaries in 
heavier DM halos retain their high values of eccentricities due
to their quick merger time $\tau_{m}$.  The connection between
$\tau_{m}$ and final eccentricity $e_{\rm LSO}$ ($e_{14}$,
$e_{22}$) can be seen even more clearly in
Fig.~\ref{fig:merger_times_VS_eLSO}, where we plot contours with
the recurrence of these PBH binary properties. We use 30
$M_{\odot}$ residing in $10^{12}\,M_{\odot}/h$ DM
halos. We have checked that allowing the PBH mass to vary
anywhere in the range $20-40\,M_{\odot}$ does not affect either
the timescale or the eccentricity results beyond the 10$\%$ level.
Observationally, LIGO and future detectors will probe a
combination of all the narrow bands of
Fig.~\ref{fig:merger_times_VS_eLSO}, since it will be difficult
to define an eccentricity at a specific pericenter distance,
given the fast evolution of the binary's orbital properties at
the late inspiral stages. 

Finally, we comment on the impact of a third PBH on the existing binaries and their properties.
A third BH traveling with a relative velocity $w'$, will affect significantly the binary if it comes 
within a distance similar to its semi-major axis $a \leq a_{0}$. The timescale for that is 
$\tau_{\rm 3rd} \sim (\pi a_{0}^{2} w' n_{\rm BH})^{-1}$, with $n_{\rm BH}$ being their local halo density. 
We compare that timescale to the binary merger timescale $\tau_{m}$. The ratio of $\tau_{m}/\tau_{\rm 3rd} \sim \pi a_{0}^{2} w' n_{\rm BH} \tau_{m}$. Using Eq.~(\ref{eq:tau_m_2}) and the results in Section~\ref{sec:Inital}, we get  $\tau_{m}/\tau_{\rm 3rd} \sim m_{\rm BH}^{3} w'^{-6} n_{\rm BH}$. Since $w' \sim (M_{\rm halo}/R_{\rm halo})^{1/2}$ and $n_{\rm BH} \sim M_{\rm halo}/V_{\rm halo}$ ($M_{\rm halo}$, $R_{\rm halo}$ and $V_{\rm halo}$ are the mass, scale radius and volume of the DM halo), we get that $\tau_{m}/\tau_{\rm 3rd} \sim N^{-2}$ where $N$ is the number of PBHs in the DM halo. We assume that even for the smallest halos $N \ge13$. Thus 3rd body interactions are negligible. 
Another possible effect includes the likely rare occasion at which a third BH of mass $m_{3}$ orbits the tight binary, causing eccentricity oscillations with a timescale $\tau_{\rm osc}$ of \cite{1997Natur.386..254H, Antonini:2015zsa}
\begin{eqnarray}
\tau_{\rm osc} = P \frac{m_{1}+m_{2}}{m_{3}} \left(\frac{a_{2}}{a_{1}} \right)^{3} (1-e_{2}^{2})^{3/2}.
\label{eq:tau_osc}
\end{eqnarray} 
$P$ is the orbital period of the wider binary (ie the binary-third BH system), $a_{1}$ the semi-major axis of the tight binary and $a_{2}$ and $e_{2}$ the equivalent orbital quantities for the binary-third BH system. Here $m_{1}$=$m_{2}$=$m_{3}$. 
Considering first
$10^{6}$ ($10^{3}$) $M_{\odot}/h$ DM halos, we take for $a_{1}$ the characteristic initial values $a_{0}$
of 3$\times 10^{5}$ ($\times 10^{3}$) au, since the slowest evolution of any binary's orbital properties occurs 
at the beginning of its existence. For $a_{2}$ we take the equivalent typical separation distance between BHs
in the halos of 2 (0.5) $\times 10^{6}$ au. Assuming also that $e_{2} \ll 1$, we get based on Eq.~(\ref{eq:tau_osc}), that $\tau_{\rm osc}/P$ = 1.5 $\times 10^{8}\, (70)$ for binaries in $10^{6}$ ($10^{3}$) $M_{\odot}/h$ DM
halos. These numbers refer to the $\tau_{\rm osc}/P$ ratio at the beginning of the tight binary's existence. As the binary evolves its semi-major axis $a_{1}$ get reduced, while the typical separation distance 
between BHs remains the same, thus the impact of the eccentricity oscillations becomes less relevant   during the evolution of the binary.
 In more massive halos the $\tau_{\rm osc}/P$ ratios are even larger. 
Furthermore, for the events that give high remaining eccentricities, the $\tau_{\rm osc}/P$ ratios are even higher than the typical values calculated above, since the relevant $a_{1}$ values are smaller in those binaries.
Thus we conclude that even if a significant fraction of binaries had a 3rd BH orbiting around, eccentricity oscillations can be ignored in this work as well.

\section{Detectability}
\label{sec:Detectability}
The detectability of an eccentric PBH-binary inspiral will depend on the exact level of the 
eccentricity before the last stable orbit (for a given noise level), as well as its distance from Earth. 
High values of eccentricities can be probed with current as well as future detectors, as we discuss next. 
We begin with an estimation of the event rate for elliptical orbits and then discuss the expected gravitational waves from these mergers.

\subsection{Eccentric event rates}
In Table~\ref{tab:TypicalFrac}, we present the calculated
fraction of 30 $M_{\odot}$ PBHs with eccentricities $e_{22}$,
$e_{14}$ and $e_{\rm LSO}$ of at least 0.1, 0.2 , 0.3 or 0.5,
respectively. We find that for PBHs in dark-matter halos of
virial mass larger than $10^{6}$ $M_{\odot}$, those fractions
are more than $O(0.01)$. Even for PBH binaries in less massive
halos the fraction is always more than 2$\times 10^{-3}$.
\begin{table*}[t]
    \begin{tabular}{cccccc}
         \hline
           Environment & $\%$ fraction & $\%$ fraction & $\%$ fraction & $\%$ fraction & $\%$ fraction \\
            $M_{\rm vir}$ ($M_{\odot}/h$) & with $e_{\rm obs}$ > 0.1 & with $e_{\rm obs} > 0.2$ & with $e_{\rm obs} > 0.3$ & with $e_{\rm obs} > 0.5$ & with $r_{p_{0}} < r_{p}$ \\
            \hline \hline
            PBHs in $10^{3}$ & 0.2, 0.4, 0.6 & 0.1, 0.2, 0.3 & 0.06, 0.12, 0.19 & 0.03, 0.06, 0.10 & 0.05, 0.15, 0.23 \\
            PBHs in $10^{6}$  & 0.6, 1.3, 2.0 & 0.3, 0.7, 1.1 & 0.2, 0.4, 0.7 & 0.1, 0.2, 0.3 & 0.3, 0.5, 1.0 \\
            PBHs in $10^{9}$ &  2.6, 5, 8 & 1.5, 2.5, 4 & 1.0, 1.6, 2.5 & 0.5, 0.8, 1.2  & 1.3, 2.2, 4  \\
            PBHs in $10^{12}$ &  8, 18, 29 & 5, 10, 15 & 3, 6, 10 & 1.5, 3, 5 & 4, 9, 15 \\
            \hline \hline 
        \end{tabular}
    \caption{The typical values for the fractions of eccentric orbits at the late stages of inspirals, 
    assuming $30\,M_{\odot}$ PBH binaries. For each case we provide three numbers, referring 
    to the eccentricities at pericenter distances $r_{p}=6$, $14$ and $22 R_{\rm Sch}$, respectively. 
    In the last column, we show the fraction of events for which the initial pericenter distance $r_{p_{0}}$ 
    is smaller than the $r_{p}$ threshold.}
    \label{tab:TypicalFrac}
\end{table*}

For LIGO we assume that eccentric events with $e_{14} > 0.3$
would be observable up to a redshift of $0.75$ at final design
sensitivity. The expected number of observed events with $e_{\rm
obs}$ up to a redshift $z_{\rm max}$ is given by
\begin{equation}
N^{e_{\rm obs}} = T_{\rm obs} \int_{0}^{z_{\rm max}} dz \; \frac{R_{m}^{e_{\rm obs}}(z) 4 \pi \chi^{2}(z)}{(1+z)H(z)},
\label{eq:fn}
\end{equation}
where $R_{m}^{e_{\rm obs}}(z)$ is the comoving merger rate of eccentric events with $e_{\rm obs}$ at the 
source redshift, $\chi(z)$ is the comoving distance and $T_{\rm
obs}$ the observing time.  Our results are 
shown in Table~\ref{tab:TypicalPar} for $e_{\rm obs} = e_{14}$.
After 6 years of LIGO final-design observations, we expect a
total of $O(1)$  events with $e_{14} > 0.2$, none of which would
originate in dark-matter halos more massive than
$10^{6}\,M_{\odot}$.  We also show forecasts for the ET, for which
the exact design sensitivity has yet to be determined. Thus, we
project for two alternative values, $z=3$ and $10$ ("optimistic"), of the
maximum redshift to detect eccentric events with $e_{14} >
0.1$. ET may observe up to a few events originating in
$10^{6}$ $M_{\odot}$ halos, and once integrating over the entire
halo mass range considered, that number increases to $O(10)$.
The contribution of the least massive halos has significant
uncertainties that originate from the properties of the DM
profile, such as its concentration, and from the discreteness of
PBH DM \cite{Bird:2016dcv}.
These uncertainties result in a wide band for the rate of PBHs that is
compatible with the updated LIGO rate mentioned in Section~\ref{sec:PBHdistr}. 
In Table~\ref{tab:TypicalPar} we used the latter for the total rate of PBHs.
For redshifts $z>6$, relevant for the optimistic ET projections, the
mass-concentration relations \cite{Ludlow:2016ifl, Prada:2011jf}
are not properly calibrated. Thus, extending them to predict rates at 
high redshifts may result in additional uncertainties.
\begin{table*}[t]
\begin{tabular}{ccccc}
\hline
Environment & $R_{m}(0)^{e_{14} > 0.2}$ & $N^{e_{14} > 0.2}$ & $N^{e_{14} > 0.1}$ & $N^{e_{14} > 0.1}$\\
$M_{\rm vir}$ ($M_{\odot}/h$) & (Gpc$^{3}$yr$^{-1}$) & LIGO 6yr & ET 10 yr & ET 10 yr (optimistic) \\
\hline \hline
PBHs in $10^{6}$ & (0.2-4) $\times 10^{-4}$ & $(0.05-1)\times 10^{-1}$ & 0.04-1 & 0.08-2 \\
PBHs in $10^{9}$ & (0.1-2.5) $\times 10^{-5}$ & $(0.2-5)\times 10^{-3}$ & $(0.2-4)\times 10^{-2}$ & $(0.5-10)\times 10^{-2}$ \\
PBHs in $10^{12}$ & (0.7-20) $\times 10^{-7}$ & $ (0.15-3)\times 10^{-5}$ & (0.25-5)$\times 10^{-3}$ & (0.04-0.8)$\times 10^{-2}$ \\
PBHs in $> 10^{2.5}$ & (1-20)$\times10^{-3}$ & 0.3-5 & 1.5-30 & 3-60 \\
BHs in GC$^{2 \rm body}$ & (0.2-2)$\times 10^{-5}$ & (1-10)$\times 10^{-3}$ & 0.1-1 & 0.3-5 \\
\hline \hline
\end{tabular}
\caption{The typical numbers of eccentric orbits at the late stages of inspiral for alternative conditions
on the host halo of $30\,M_{\odot}$ PBHs. The rate quoted is the comoving merger rate. In the last three
columns we present the number of events we expect LIGO (final design) or ET would observe in a 6 or 10 year
interval, respectively. The values in the last column account for optimistic alternative assumptions
regarding the ET sensitivity (see text for details). For the case where we include all PBHs in halos of
$10^{2.5}\,M_{\odot}/h$ or heavier (4th row), we give a range in accordance to the LIGO quoted rate.
The relative contribution from different DM halos is based on the results of \cite{Bird:2016dcv}.
In the bottom row we show the expected contribution from 2-body encounters of BHs in the cores of
globular clusters, with uncertainty ranges reflecting the major astrophysical uncertainties (see text for details).}
\label{tab:TypicalPar}
\end{table*}
We note that if instead of the fixed value $30\,M_{\odot}$ for the PBHs, their mass is allowed to be in the wider 
range of $20-40\,M_{\odot}$, the number of eccentric coalescence events presented in Table~\ref{tab:TypicalPar} 
changes by only $\simeq 10 \%$, increasing with decreasing BH mass. 

\subsection{Alternative models to PBHs}

As mentioned in the Introduction, there are a number of
alternative models for the progenitors of BH binaries whose
mergers may be seen in gravitational-wave observatories.
For example, globular clusters (GCs) may provide
fertile breeding grounds for binary black holes. Most binaries
in GCs are formed by 3-body interactions, where the BHs form a
binary that becomes tighter after multiple interactions with
nearby BHs over a time span of Gyrs. Moreover, through those
same interactions the binaries receive random kicks, and
eventually get ejected from their host GC
\cite{Morscher:2012se,Chatterjee:2016hxc}, after which they
evolve as isolated systems. As a consequence, these systems are
typically highly circularized well before their GW emission
enters the LIGO sensitivity band
\cite{Rodriguez:2016kxx,Harry:2016ijz}.  Yet, there is a small
fraction of BH binaries formed through 2-body interactions at
the cores of GCs, which may retain their eccentricities up to
the last stages of the inspiral. We study these binaries, and
refer to them as ``BHs in GC$^{2 \rm body}$" (fifth row in
Table~\ref{tab:TypicalPar}).

A representative distribution of the structural parameters of
GCs can be found on the Harris catalog \cite{Harris:1996kt}. We
focus on the objects of this catalog with a measurement of their
core velocity-dispersion.  For these GCs we can also retrieve
their core radii, distributed log-normally around 1 pc, and
their core luminosities. We calculate the GC core mass from its
luminosity, assuming a light-to-mass ratio of
$1\,L_{\odot}/M_{\odot}$.  Given these three parameters, we can
calculate the number of 2-body captures that would occur in a GC
core \cite{O'Leary:2008xt}, assuming that BHs make up all its
mass. We find this binary formation rate to be between $\sim
10^{-10}$ and $10^{-11}$ yr$^{-1}$ per GC, for GCs with core
radii in the range from 0.1 pc to 1 pc. Smaller cores would not
have enough BHs to have any meaningful interactions, and bigger
cores would be too diffuse.

However, not all the core mass will be in the form of BHs. With
the Kroupa initial mass function
\cite{Kroupa:2000iv,Rodriguez:2016kxx}, only $1\%$
of the mass of the entire GC is in objects 
heavier than 30 $M_{\odot}$. The core mass is around one tenth
of the total GC mass; thus yielding an upper estimate of $f_{\rm
BH} \simeq 0.1$ to the fraction of the core mass in  30
$M_{\odot}$ BHs (that is if all the 30 $M_{\odot}$ BHs of a GC
are concentrated in its core).  Knowing $f_{\rm BH}$, and since
there are roughly $0.7$ GCs per ${\rm Mpc}^3$
\cite{Rodriguez:2015oxa}, we obtain a local merger rate of
$R_{m}(z=0)$= $7 \times 10^{-2}-7\times 10^{-3}$ $\times f_{\rm
BH}^{2}$ Gpc$^{-3}$ yr$^{-1}$.  We note that this rate is a realistic upper
bound. Many effects can change it. Many BHs from
stars more massive than 30 $M_{\odot}$ end up becoming lighter
BHs, which should not be confused with the PBHs that we are
after.  Moreover, it has been suggested that heavy BHs are
expelled from GCs early on \cite{Rodriguez:2016kxx}.
Alternatively, dynamical friction can bring the velocity dispersion of
the BHs below that of the stars in the GC core, decreasing their
virial radius and making them more concentrated, hence producing
more events.

Of course, not all the BH mergers due to 2-body interactions in
the cores of GCs will be highly eccentric.  To find the number
of these events we have assumed that the BHs follow a
Maxwell-Boltzmann relative velocity distribution with $v_{\rm
DM} = 12$ km~s$^{-1}$ \cite{Harris:1996kt}.  Thus, while GCs can in
fact be the major source of GW detections, coming from circular
BH binaries, we do not expect LIGO to observe any heavy binary
mergers with high eccentricities (we estimate less than $O(10^{-2})$
events). With ET, and after 10 yrs of observations, 
there is an optimistic estimate of $O(1)$ eccentric events from
GCs; to be compared to the conservative $O(10)$ events from
PBHs. We think that these two alternatives remain
distinguishable. 

The authors of Ref.~\cite{O'Leary:2008xt} have suggested another
scenario for binary BHs with eccentric inspirals in the LIGO
frequency range. These binaries reside in the inner sub-pc of
galactic nuclei and are formed by 2-body interactions of BHs
in a gravitational potential dominated by
the central supermassive BH. Such black hole binaries, with relative
velocities of $\simeq$ 30-80 km~s$^{-1}$, would also be
formed with high initial eccentricities and relatively small
merger timescales. We get that $\simeq 10\%$ of such binaries would 
retain a high eccentricity. However, the model of Ref.~\cite{O'Leary:2008xt} 
relies on densities of BHs that are extremely uncertain in the inner sub-pc of the 
galactic nuclei, with those densities varying by typically $\sim$ 10 orders of magnitude 
(20 orders of magnitude for the rate profile) from $10^{-4}$ to 1 pc.
In particular, using the models
"B" and "E-2" of \cite{O'Leary:2008xt} for which a density profile is given,
we find that $\simeq 60 \%$ ( $\simeq 95 \%$) of the total merger rates
are due to $O(10)$ ($O(10^{2})$) BHs in the inner $10^{-3}$ ($10^{-2}$) pc.

In our opinion, all these uncertainties are simply too large to allow a fair
comparison to the PBHs case. Yet, we note that if one takes their
assumptions at face value, then LIGO should be able to detect anywhere between
a few to several hundreds of eccentric events during its expected
observation period, with a factor of 10 more total (nearly circularized) events.
Yet, all those events should still follow a mass function, originating from the
initial stellar mass-function, with the majority of those events being
5-10 $M_{\odot}$ merging BHs.
That is in stark contrast to the  narrower and higher mass range of the PBHs 
considered here.

\subsection{Waves from  eccentric inspirals}
\label{sec:HighModesObservations}

In the previous Section we discussed the properties of PBH
binaries and the rate of occurrence of gravitational-wave
inspirals with high eccentricities, showing that with fiducial
assumptions we could perhaps observe such events with LIGO. In
this Section, we discuss some observational aspects of these  
high-eccentricity events. 

Every coalescence event has an inspiral phase lasting
$\tau_{m}$, a merger phase, and a ringdown phase.  Each
observatory can, through the portion of the inspiral phase that
is within its frequency range, measure the redshifted chirp
mass, which at the position of the source is
\begin{eqnarray}
M_{c} = \frac{(m_{1} \cdot m_{2})^{3/5}}{m_{\rm tot}^{1/5}}.
\label{eq:Mchirp}
\end{eqnarray}
Following Ref.~\cite{Flanagan:1997sx}, the spectral energy
density at the source of the emitted GWs during the 
inspiral of a circularized orbit is
\begin{eqnarray}
\frac{dE}{df_{s}}_{\rm inspiral} = \frac{1}{3} \left( \frac{\pi^{2}}{f_{s}} \right)^{1/3} 
\frac{m_{2} \cdot m_{2}}{m_{\rm tot}^{1/3}},
\label{eq:dEdf_Insp}
\end{eqnarray}
where $f_{s}$ is the GW frequency at the source ($f_{\rm obs} = f_{s}/(1+z)$).
The frequency at the end of the inspiral and the beginning of
the merger phase is (at the source),
\begin{eqnarray}
f_{\rm merger} (m_{1}, m_{2}) = 0.02/m_{\rm tot}.
\label{eq:fmerg}
\end{eqnarray}
Between the redshifted $f_{\rm merger}$ and the frequency of
quasi-normal ringdown (at the position of the binary),
\begin{eqnarray}
f_{\rm ringdown} (m_{1}, m_{2}) = \frac{
\left(1-0.63(1-\alpha)^{3/10} \right)}{2 \pi m_{\rm tot}},
\label{eq:fqnr}
\end{eqnarray}
the merger phase of the coalescence events is observed. 
The dimensionless spin $\alpha$ of the final BH is simply
$\alpha = \frac{c S}{G m_{f}^{2}}$, assuming $m_{f} \simeq
m_{\rm tot}$. The merger phase is taken to last
\begin{eqnarray}
\tau_{\rm merger} (m_{1}, m_{2}) = 14.7 \frac{m_{\rm tot}}{10^{5} M_{\odot}} \textrm{s}.
\label{eq:taum}
\end{eqnarray}

During the merger phase, the spectral energy density is given by
\begin{eqnarray}
\frac{dE}{df_{s}}_{\rm merger}= \frac{16 \mu^{2} \epsilon}
{m_{\rm tot}(f_{\rm ringdown} - f_{\rm merger})},
\label{eq:dEdf_Merg}
\end{eqnarray}
where $\mu$ is the reduced mass and $\epsilon$ is the fraction
of the energy in the initial BH binary that is emitted in GWs
during the merger phase. We take $\epsilon = 0.04$, in agreement
with the uncertainties of the GW150914 event
\cite{TheLIGOScientific:2016wfe}.  Alternative values for
$\epsilon$ and $\alpha$ will only affect the merger and ringdown
phases, which are of no direct importance for the eccentricity
discussion here. 

The observed strain amplitude is
\begin{eqnarray}
h_{c}(f_{\rm obs})  = \sqrt{2}\frac{1+z}{\pi d_{L}(z)}\sqrt{\frac{dE}{df_{s}}},
\label{eq:hc}
\end{eqnarray}
where for the $\frac{dE}{df_{s}}$ we include the inspiral and
merger phases, but ignore the contribution from the ringdown,
which is short and characterized by a very fast reduction of the
$h_{c}$ with time (even in the high signal-to-noise GW150914
event).

\begin{figure*}
\begin{centering}
\includegraphics[width=4.2in,angle=0]{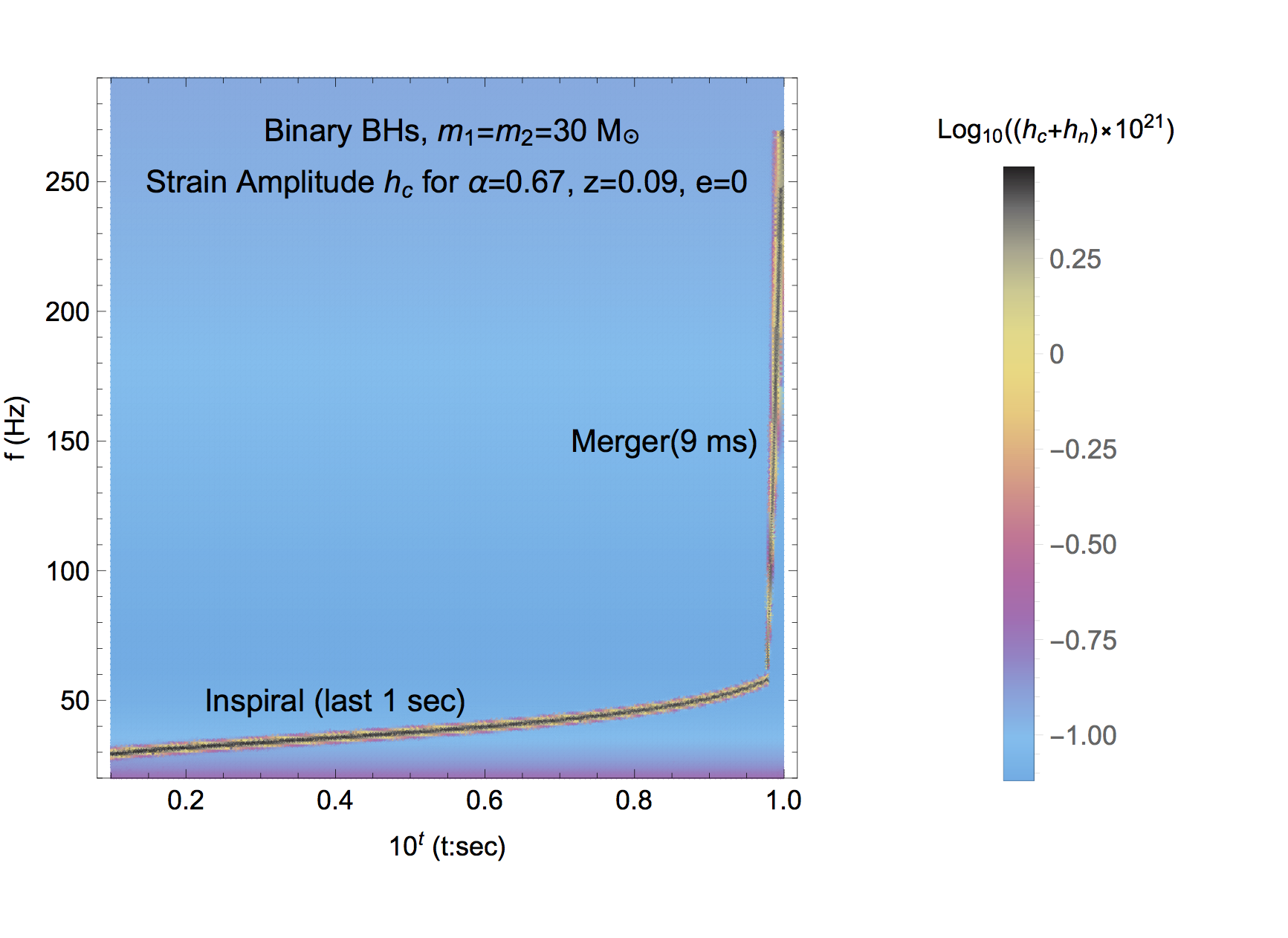}
\hspace{-3.7cm}
\includegraphics[width=4.2in,angle=0]{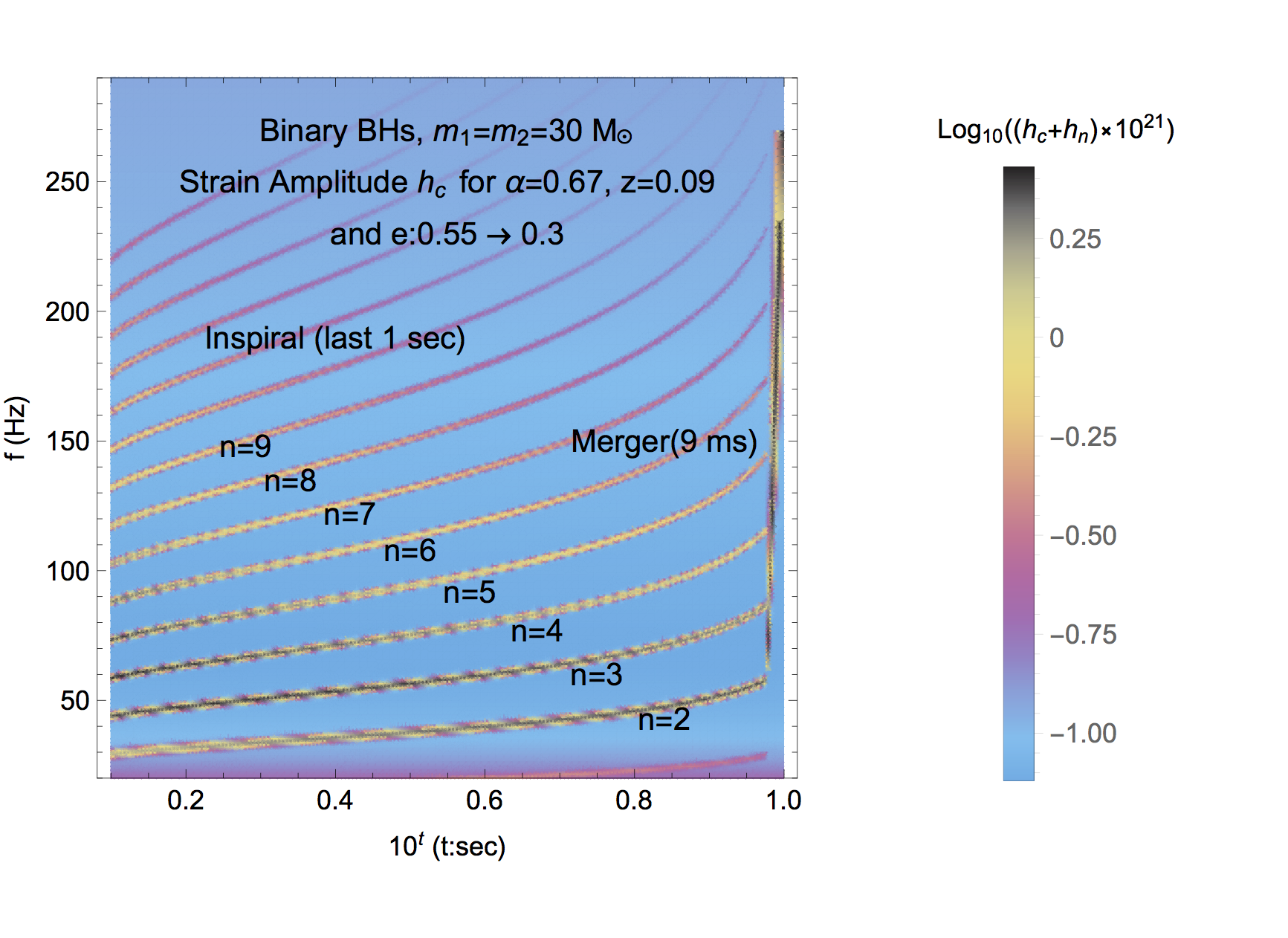}
\end{centering}
\caption{Evolution over the last second of the amplitude of the signal strain, superimposed 
on the noise as expected with the LIGO design sensitivity. Time $t$ here is between -1 to 0 seconds. \textit{Left}: assuming no significant 
remaining eccentricity over that last one second. \textit{Right}: assuming that during the last 
one second there is a remaining eccentricity that evolves from 0.55 to 0.3, resulting in the 
presence of higher GW modes than just the quadrupole (n=2) on. The different mode amplitudes
also evolve with time.}
\label{fig:hc_vs_time_LIGO}
\end{figure*}

In Fig.~\ref{fig:hc_vs_time_LIGO} (\textit{left} panel), we show
the evolution of the strain amplitude over frequency and time
during the last second of the coalescence of two 30 $M_{\odot}$
BHs following a circularized orbit. 
Assuming that the event occurs at a redshift of 0.09 and that
the resulting final BH has a spin $\alpha$ of 0.67 (roughly
corresponding to the best-fit values for the GW150914 event), we
see that such an event would be easily traced over the expected
final design noise of LIGO \cite{Aasi:2013wya}, during a full
second before the merger.  If instead, those black holes were on
an elliptical orbit with eccentricity $e$ evolving from 0.55 to
0.3 during that last second of the coalescence (\textit{right}
panel of Fig.~\ref{fig:hc_vs_time_LIGO}), then GW power would be
emitted also in other modes that at a given time are emitted
from the source at frequencies $f_{n_{\rm source}} = n \cdot f_{\rm
orb}$, where $f_{\rm orb}$ is the Keplerian orbital frequency of
the binary.  As can be clearly seen, in addition to observing
the $n=2$ (quadrupole) mode, LIGO with its expected final-design
sensitivity should clearly be able to identify higher modes at least up
to $n=8$, since for frequencies $> 50$Hz all these additional
modes have a strain amplitude that is at least a factor of 3
higher than that of the noise.  We discuss the details of how
the strain amplitudes for those higher modes are calculated in
Appendix~\ref{app1}.  We note that the extent to which the
higher modes can be identified relies on the waveforms  
used by the LIGO collaboration. We also note the eccentricity 
gets reduced within the last second of the inspiral, which
changes the relative power of GWs between modes. 
As the eccentricity of the orbit is reduced, lower modes down to the quadrupole,
become more powerful over higher ones \cite{Peters:1963ux}. That
should also be a matter of further investigation, to be
accounted for by the waveform searches. That is because the rate
of eccentricity evolution with time over the last phase of the
inspiral, depends on the exact realization of the BH binary
coalescence.

\begin{figure*}
\begin{centering}
\includegraphics[width=4.2in,angle=0]{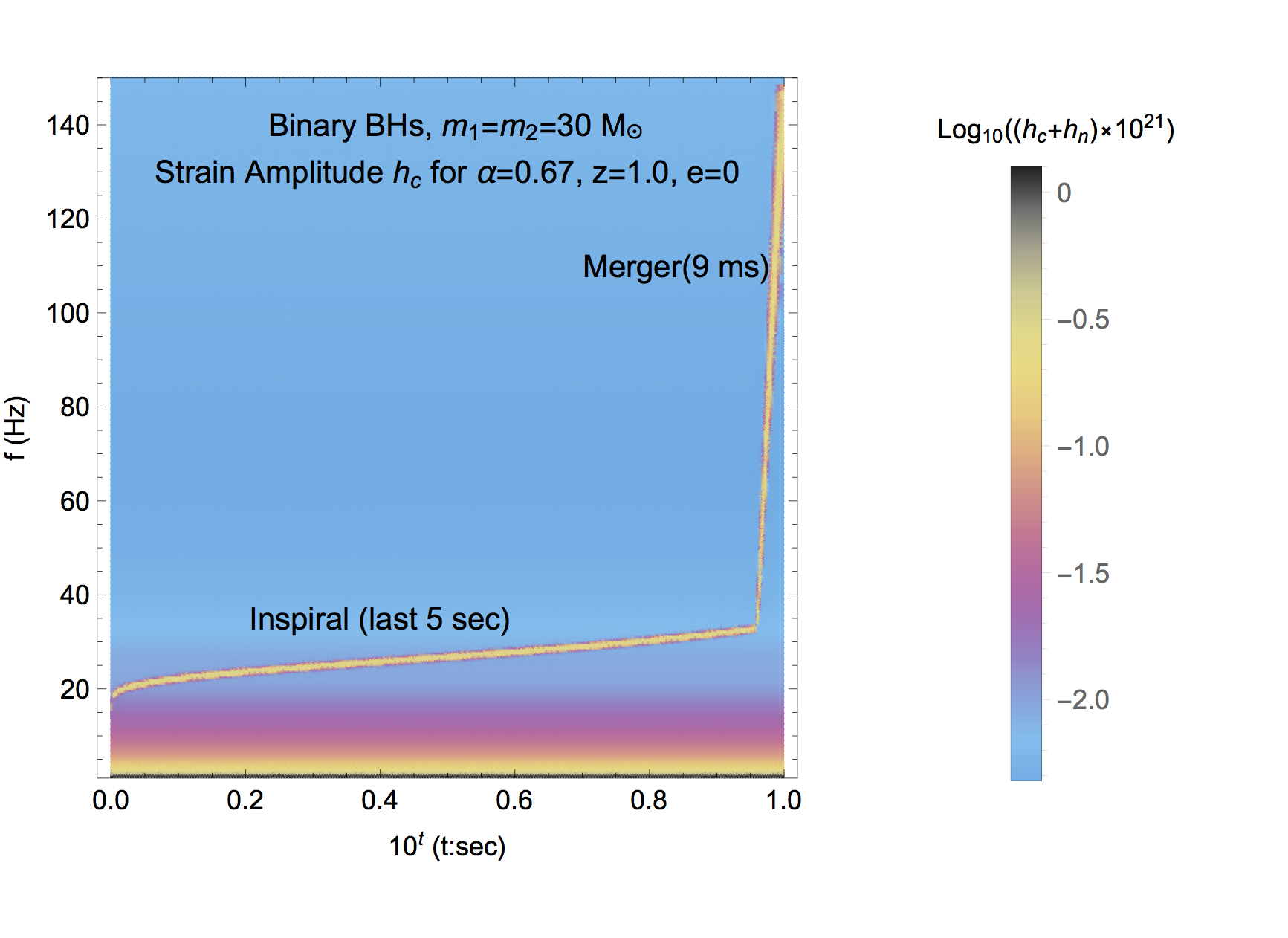}
\hspace{-3.7cm}
\includegraphics[width=4.2in,angle=0]{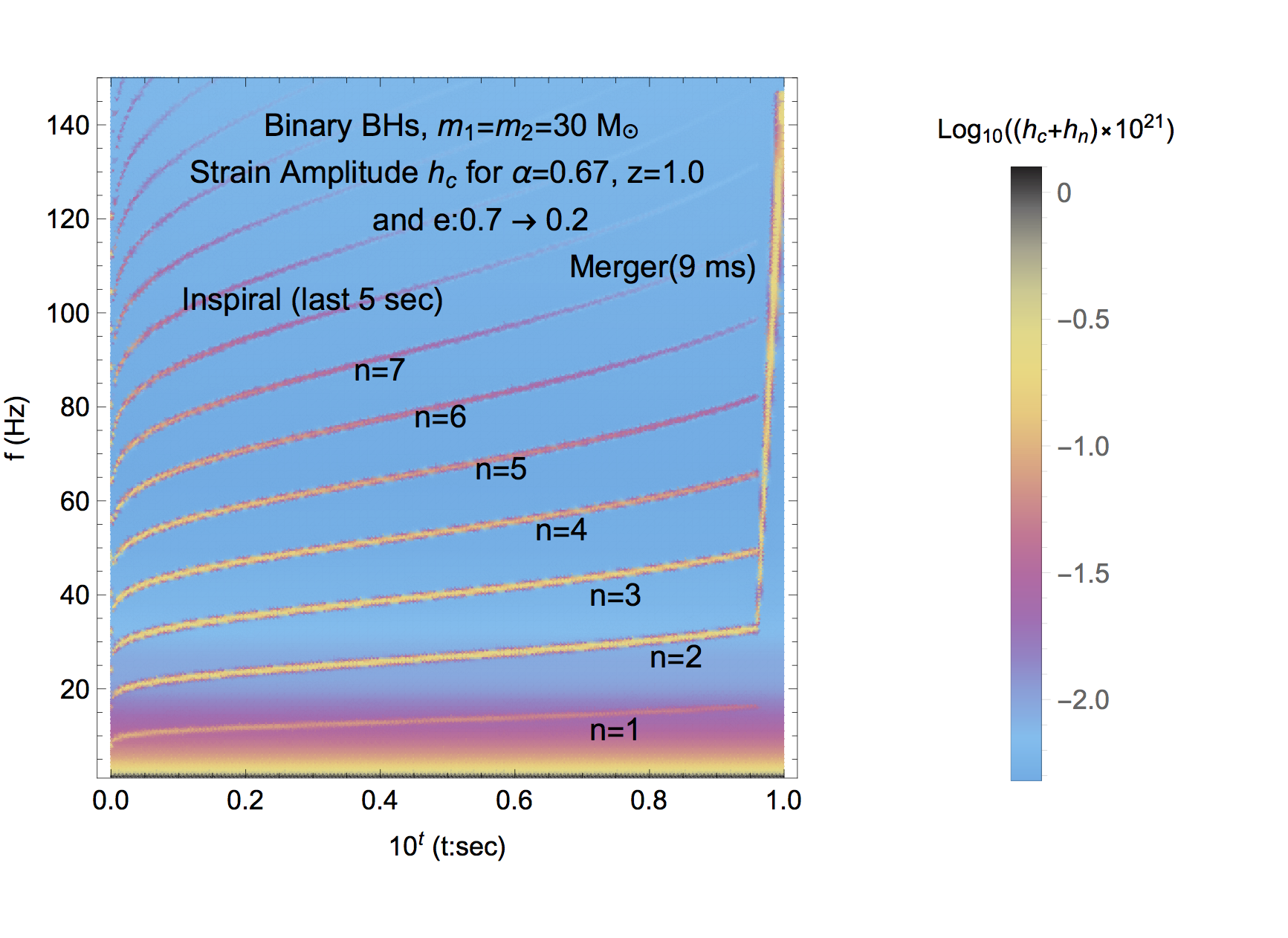}
\end{centering}
\caption{As in Fig.~\ref{fig:hc_vs_time_LIGO}, assuming ET noise sensitivity instead. We plot the 
last 5 seconds. \textit{Left}: assuming no significant remaining eccentricity over that last 5 seconds. 
\textit{Right}: assuming that during the last five seconds there is an evolving eccentricity from 0.7 to 0.2.
One can also see an $n=1$ mode just above the background.}
\label{fig:hc_vs_time_ET}
\end{figure*}

With future detectors such as ET, the significantly lower noise
and wider frequency range will allow to follow coalescence
events from further away and for a longer time.
Fig.~\ref{fig:hc_vs_time_ET} shows an equivalent merging event
of two 30 $M_{\odot}$ BHs, but at a redshift of 1 and with an
eccentricity of $e = 0$ ($e : 0.7 \rightarrow 0.2$) \textit{left} (\textit{right}), as that could be observed by ET.  
The first several higher modes could be followed for several seconds, possibly allowing even to probe 
the time evolution of the eccentricity of the individual coalescence events.  

Observing higher modes can also allow the identification of
events with eccentricity at the last stages of the 
inspiral with a higher signal-to-noise ratio (S/N) than that of the equivalent circularized objects. 
In fact, properly accounting for the presence of higher modes is relevant to understanding the physical 
properties (mainly the masses and the distance) of the binary.
In Table~\ref{tab:StoN}, we present the expected S/N at LIGO and ET. For LIGO we assume the final 
design sensitivity while for the Einstein Telescope we used the
design option "ET-B" of Ref.~\cite{Sathyaprakash:2012jk}, 
which is the more pessimistic at low frequencies (relevant for high-mass BH coalescence events). 
As can be seen, the exact contribution to the S/N from the various GW modes at the inspiral 
depends on the eccentricity of the binary once its GWs enter the frequency band of the observatories, 
denoted here as $e_{\rm in}$ (not to be confused with the eccentricity at formation of the binary $e_{0}$),
as well as the final $e_{\rm LSO}$. Yet, for events with significant eccentricities during observation, 
higher modes can contribute significantly to the total
S/N. Furthermore, these modes would reduce 
the overall contribution of the quadrupole mode to the S/N.

\begin{table*}
    \begin{tabular}{cccccccccc}
         \hline
           Observatory & z & $e_{\rm in}$$\rightarrow$$e_{\rm LSO}$  & S/N Merger & $\Delta$(S/N) Ins. & $\Delta$(S/N) Ins. & $\Delta$(S/N) Ins. & $\Delta$(S/N) Ins. & $\Delta$(S/N) Ins. & $\Delta$(S/N) Ins. \\
           final design  &  &  & $\&$ Ringdown & (n=2) & (n=3) & (n=4) & (n=5) & ($6 \leq$n$\leq 10$) & ($11 \leq$n$\leq 15$) \\
            \hline \hline
            LIGO  & 0.09 & 0 $\rightarrow$ 0 & 44 & +25 & - & - & - & - & - \\
            LIGO  & 0.09 & 0.6 $\rightarrow$ 0 & 44 & +22 & +33 & +35 & +31 & +85 & +19 \\
            LIGO  & 0.09 & 0.6 $\rightarrow$ 0.3 & 44 & +8.0 & +30 & +37 & +33 & +89 & +19 \\
            LIGO  & 0.75 & 0 $\rightarrow$ 0 & 6.8 & +1.3 & - & - & - & - & - \\
            LIGO  & 0.75 & 0.6 $\rightarrow$ 0.3 & 6.8 & +1.2 & +3.3 & +4.2 & +4.1 & +11.8 & +2.6 \\
            ET  & 1 & 0 $\rightarrow$ 0 & 82 & +16 & - & - & - & - & - \\
            ET  & 1 & 0.6 $\rightarrow$ 0.2 & 82 & +8.1 & +35 & +49 & +50 & +145 & +32 \\
            \hline \hline 
        \end{tabular}
    \caption{The contribution of higher GW modes at elliptical orbits of 30 $M_{\odot}$ PBHs. 
    We assume that when entering the 
    frequency band of observation the binary has an initial eccentricity $e_{\rm in}$ that evolves down 
    to $e_{\rm LSO}$. The total S/N is the linear sum of columns 4-10 (we have accounted for the quadratic 
    sum from the various phases). We take $\alpha = 0.67$ in Eq.~(\ref{eq:fqnr}) and $\epsilon = 0.04$ in Eq.~(\ref{eq:dEdf_Merg}). If we let $\alpha : (0.5-0.9)$ and $\epsilon : (0.03-0.05)$, there is a resulting 20$\%$ uncertainty in the values quoted in the fourth column.}
    \label{tab:StoN}
\end{table*}

The results in Table~\ref{tab:StoN} do not include the $n=1$ mode since 
its contribution to the S/N, is small and also depends on the exact assumptions of the instrument
sensitivity at the lowest frequencies.  
For LIGO, the frequency band we assume starts as 20 Hz and for ET that frequency is 10 Hz. 
Any further advances that would allow those conservative values to be reduced would increase the 
S/N values quoted and the capacity of these observatories to identify elliptical orbits at coalescence. 

The last orbits of the inspiral, the merger, and the ringdown,
have been the subject of numerical relativity and a variety of
analytic approaches, where different
orders of post-Newtonian approximation are implemented 
and where the impact of BH spins are included 
\cite{Buonanno:2002fy, Blanchet:2004ek, Blanchet:2006gy, Blanchet:2008je, Arun:2008kb, Ajith:2009bn, Ajith:2011ec}. 
While future advances will certainly help in identifying GW events with higher modes, our quantitative results in this 
Section should not be affected by these future developments. Such developments will affect predominantly 
the contributions of the merger and inspiral phases to the total S/N (fourth column in Table~\ref{tab:StoN}). 
For the inspirals, more accurate waveforms in combination with lower instrumental noise will only affect 
how many of the higher modes will be identified and contribute to the S/N. 

\section{Conclusions}
\label{sec:Conclusions}

The possibility that the merger of primordial black holes may
provide at least some of the gravitational wave signals to be
seen in the coming decade, was first discussed in
Ref.~\cite{Bird:2016dcv}. In this work, we investigated some of the
implications of that hypothesis for future observations of
gravitational waves.

We found that all binaries of PBHs, in all DM halos, are formed
with high eccentricities. In the more massive halos, these
binaries are characterized by small separations and short merger
timescales, which can be as long as years and in some cases less
than minutes.  Consequently, some of the binaries
retain a significant portion of their initial eccentricity as
they enter the frequency bands of GW detectors. Even in the
smallest halos we considered, $\sim 10^{3} M_{\odot}$, there is
a $O(1)\%$ chance that a merger will have a high
eccentricity.  Altogether, advanced LIGO is expected to observe
$O(1)$ events with large eccentricities at the final stages of
inspiral, while the ET should observe $O(10)$ such 
events. 

These coalescences are characterized by detectably strong
higher modes of GW waves. In fact, with ET we should be able to easily 
observe multiple higher modes from PBH binaries as far as
redshifts $z \gsim 1$.  The development of strategies to search
for these higher modes is important, since misinterpretation of
the physical properties (masses and redshift) of the coalescing
binaries could result if they were missed. In addition, the total S/N
can be significantly increased by measuring these higher modes.

We also considered eccentric events from competing astrophysical
sources and found them likely to be a factor of 10 smaller, although
uncertainties prevent a more definitive statement.   We
therefore conclude that a detection of highly-eccentric
$30\,M_{\odot}$ GW events will provide strong supporting
evidence for the PBH progenitor scenario. Since for some of the PBH 
binaries, the merger timescale is $\sim$ years or less, with future 
observations at lower frequencies, such as those by eLISA \cite{AmaroSeoane:2012je},
we may be able to follow such binaries through a longer 
era of their evolution. In fact, if proper waveforms are developed we may 
be able to detect GW emission from the formation event. 

Strong evidence for
a PBH-merger contribution to gravitational-wave events will
likely require additional lines of evidence, including possibly
some from cross-correlations of high-mass GW event
locations with galaxy surveys \cite{Raccanelli:2016cud}, or an
excess in the stochastic gravitational-wave background at
lower frequencies (from higher redshifts, where PBH
should still reside) \cite{GWBpaper}.  Furthermore, PBH binaries are
formed from randomly moving BHs. We thus expect no correlation
between the products $\vec{S_{1}}\cdot \vec{L}$,
$\vec{S_{2}}\cdot \vec{L}$, where $\vec{S_{i}}$ is the spin of
the individual BHs and $\vec{L}$ the angular momentum (note that
this is a characteristic that is shared for all binaries formed
through dynamical processes, rather than a common-envelope
origin).

Ultimately, if the scenario considered here is correct,
the mass distribution of an ensemble of events should
demonstrate a significant contribution from high-mass BHs, and
their measured properties, such as the eccentricity and spin,
should be consistent with the two-body capture mechanism. In
Ref.~\cite{Bird:2016dcv} we estimated that by the end of a
six-year run at full  sensitivity, LIGO should observe $\sim
600$ PBH events. ET will
observe roughly an order of magnitude more events. This should
provide ample statistics to perform these tests.
                  
\begin{acknowledgments}
                  
We would like to thank David Kaplan and Vuk Mandic for interesting discussions.
SB was supported by NASA through Einstein Postdoctoral
Fellowship Award Number PF5-160133. This work was supported by
NSF Grant No. 0244990, NASA NNX15AB18G, the John Templeton
Foundation, and the Simons Foundation.
\end{acknowledgments}

\bibliography{GW_elliptical}
\bibliographystyle{apsrev}

\begin{appendix}

\section{Strain amplitude for higher modes}
\label{app1}

Here we provide formulas needed to calculate the strain
amplitude $h_{c}$ used in Section~\ref{sec:HighModesObservations}.

We define $f_{\rm orb}^{\rm min}$ at the minimal orbital
frequency where there is contribution to GWs at detectable
frequency ranges by a given experiment. GW modes are emitted at
frequencies of 
\begin{equation}
f_{n} = n \frac{f_{\rm orb}}{1+z} \; \textrm{with} \; n \geq 2. 
\label{eq:fn}
\end{equation}
The current sensitivity of LIGO has a minimum of $f^{\rm min} =
20$ Hz with the design sensitivity being $f^{\rm min} = 10$ Hz,
and with the future ET expected design sensitivity being at
$f^{\rm min} = 2$ Hz. Thus the relevant $f_{\rm orb}^{\rm min}$
are 10, 5 and 1 Hz respectively. In this work we use $f_{\rm
orb}^{\rm min} = 10$ Hz for LIGO and 5 Hz for ET as a
conservative estimate.  Thus, for a given eccentricity, LIGO or
ET observe the inspiraling binary when its pericenter is at most
\begin{eqnarray}
r_{\rm per}^{\rm in}(m_{1}, m_{2}, e) = \left(\frac{2 \pi}{f_{\rm orb}^{\rm min}}  (1+e^{\rm in})^{1/2} G^{1/2} m_{\rm tot}^{1/2} \right)^{\frac{2}{3}},~~~~~~
\label{eq:rper0}
\end{eqnarray}
or in its dimensionless form at
\begin{eqnarray}
\rho(m_{1}, m_{2}, e^{\rm in}) = \frac{r_{\rm per}^{\rm in}(m_{1}, m_{2}, e^{\rm in}) G^{-1}c^{2}}{m_{\rm tot}}. 
\label{eq:rho}
\end{eqnarray}

The spectral energy densities of higher modes at a given
eccentricity, can be evaluated based on the quadrupole mode at
the same time by \cite{Peters:1963ux}
\begin{eqnarray}
\frac{dE}{df_{s}}_{\rm inspiral}^{n \geq 2} = \frac{2}{n} \frac{g(n,e)}{g(2,e)} \frac{dE}{df_{s}}_{\rm inspiral}^{n=2},
\label{eq:dEdf_Insp}
\end{eqnarray}
where
\begin{eqnarray}
g(n,e) = \frac{n^{4}}{32} 
&&\left[  \left( J_{n-2} - 2 e J_{n-1} + \frac{2}{n} J_{n} + 2eJ_{n+1} - J_{n+2}  \right)^{2} \right. \nonumber \\
&& + \left.(1-e^{2})\left( J_{n-2} - 2J_{n} +J_{n+2} \right)^{2} + \frac{4}{3 n^{2}} J_{n}^{2} \right], \nonumber \\
\label{eq:g}
\end{eqnarray}
and $J_{n}$ is the Bessel function of order $n$ evaluated at $ne$. For $n=1$ we use that \cite{Peters:1963ux}
\begin{eqnarray}
\sum_{n=1}^{n = \infty} = \frac{1+ \frac{73}{24}e^{2} + \frac{37}{96}e^{4}}{(1-e^{2})^{7/2}}.
\label{eq:n_eq_1}
\end{eqnarray}
For the total signal-to-noise ratio we used \cite{O'Leary:2008xt}
\begin{eqnarray}
\langle S^{2}/N^{2}\rangle_{\rm inspiral} &=& 2\frac{48}{95} \frac{\eta (m_{\rm tot} (1+z))^{3} \rho(m_{1}, m_{2}, e^{\rm in})}{dL^{2}(z)} \nonumber \\
&&\sum_{n=1}^{n_{\rm max}} \int_{f^{\rm min}}^{f^{\rm max}} \frac{df}{f} \frac{n \, g(n,e) \, s(e, e_{\rm in})}{e \, (1+z) \, f \, S_{n}(f)}
\label{eq:SN_insp}
\end{eqnarray}
and
\begin{equation}
\langle S^{2}/N^{2}\rangle_{{\rm merger} + {\rm ringdown}} = 2\frac{4}{5} \int_{f^{\rm min}}^{f^{\rm max}} 
df \frac{h_{c}^{2}(f)}{S_{n}(f) (2 \, f)^{2}}
\label{eq:SN_MergQNR}
\end{equation}
with
\begin{eqnarray}
s(e,e_{\rm in}) &=& \left( \frac{e}{e_{\rm in}} \right)^{\frac{24}{19}} \left( \frac{1 + \frac{121}{304} e^{2}}{1 + \frac{121}{304} e_{\rm in}^{2}} \right)^{\frac{1740}{2299}} \times\nonumber \\
&&\times\left(\frac{(1+e_{\rm in}^{2}) (1 - e^{2})^{3/2}}{1 - \frac{183}{304}e^{2} - \frac{121}{304} e^{4}} \right).
\label{eq:fn}
\end{eqnarray}
In Eq.~(\ref{eq:SN_insp}) and~(\ref{eq:SN_MergQNR}), $S_{n}(f)$
is the strain noise amplitude ($h_{n}(f)$) squared, f  is
$f_{\rm obs}$.

\end{appendix}

\end{document}